\documentclass[aps,pre,twocolumn,groupedaddress,showpacs]{revtex4}
\usepackage{graphicx}
\usepackage{epstopdf}
\begin{document}


\title{Shell Inhomogeneities of the Single-Particle Level Spectra in the Practical Model of the Gamma-Decay of Neutron Resonance
}
\author{Jovan\v cevi\' c N.$^{1}$, Mitsyna L.V.$^{2}$, Sukhovoj A.M.$^{2}$, Kne\v zevi\' c D.$^{1,4}$, Krmar M.$^{1}$, Petrovi\'c J.$^{1}$,  Oberstedt S.$^{3}$, Dragi\' c A.$^{4}$, Hambsch F.-J.$^{3}$ and  Vu D.C.$^{2,5}$ }
\affiliation{$^1$ University of Novi Sad, Facility of Science, Department of Physics, Trg Dositeja Obradovica 3, 21000 Novi Sad, Serbia,}
\affiliation{$^2$ Joint Institute for Nuclear Research, Dubna, 141980, Russia}
\affiliation{$^3$ European Commission, Joint Research Centre, Institute for Reference Materials and Measurements (IRMM), Retieseweg 111, B-2440 Geel, Belgium}
\affiliation{$^4$ Institute of Physics Belgrade, Pregrevica 118, 11080 Zemun, Serbia}
\affiliation{$^5$ Vietnam Atomic Energy Institute, Vietnam}
\date{\today}


\begin{abstract}
In this paper we have taken into account shell inhomogeneities of the single-particle level spectra in the practical model of the gamma-decay of neutron resonance developed in Dubna. The obtained data confirm a dependence of breaking thresholds for Cooper pairs on a shape of investigated nuclei -- a phenomenon noticed earlier without taking into account the shell inhomogeneity.

\end{abstract}
\maketitle

\section{Introduction}
\label{intro}

The components of the wave function for each nuclear level increase in number with the growth of the excitation energy, while the absolute values of these components have a decrease \cite{Solo,Solo1}. This effect is explained in the theory of the nucleus, in the framework of the quasi-phonon model \cite{Solo2,Weid,Iach}. Undoubtedly, this effect plays the leading role when investigating the behavior of nuclear matter. 

All experiments that investigate the structure of an excited nucleus are based on measuring the spectra i.e. the cross sections. For complete and reliable studies of gamma--decay processes, the experiment has to allow for the obtaining of both the level density, $\rho$, and the emission widths for products of the nuclear reaction $\Gamma$, for all excited levels, from the measured intensity of the spectra \cite{Igna}. The main problem for obtaining the values of $\rho$ and $\Gamma$ for all decaying levels is the absence of spectrometers with an energy resolution $FWHM < D_i$ (for any spacing $D_i$ between intermediate levels), as well as picosecond time resolution for any excitation energy of investigated nucleus. Because of these limitations only an average number of excited levels and an average value of partial widths in fixed energy intervals can be measured. 
\subsection{Methods for determination of nuclear parameters}
\label{model}

Nuclear parameters extracted from the measured spectra describe the process of emission of the reaction products. For studying the nuclear structure there are two different procedures ways usually named as "one-step" \cite{Vona,Zhur,Barth,Schi,Lars1,Lars2} and "two-step" reactions \cite{Valn,Honz,Sukh,Popov,Bonda,Bonev,Vasi,Sukh1}. 

In the case of the one-step reaction, any gamma-quanta (or nucleon) of the compound-state decay is recorded irrespective of the energy of the excited level (the total energy of all reaction products is equal to the compound-state excitation energy). In the two-step reaction, a coincidence of two gamma-quanta of the same cascade is recorded. At that, the secondary gamma-transitions of the cascade to a group of low-lying levels (including the ground state of the nucleus) are also recorded. Only in the two-step experiment the information about the energy of intermediate level is included in the data treatment process. 

The fundamental difference between the two-step and one-step experiments becomes evident, when the level density is obtained from the evaporated spectra.  As the correlation of the level density and penetrability coefficients on the wave function of excited level are not taken into account in the one-step experiment, only the product of $\rho$ and $\Gamma$ functions can be determined. Additionally, because of the strong anti-correlation of the $\rho$ and $\Gamma$ functions, an unknown systematic uncertainty of their determination appears. 

Only in two-step experiments there is a possibility to reduce the uncertainty of the $\rho$ and $\Gamma$ determination, as they are described by appreciably different functions. A decrease in the methodical errors is due to the fact that the intensity $I_{\gamma \gamma}\left(E_1\right)$ as a function of primary transition energy $E_1$ is, in essence, a convolution of two practically independent experiments, i.e. we can consider the spectrum of primary transitions and the branching ratio distribution of secondary transitions as independent distributions. 

A rise in the quality of the data of the two-step experiments is obtained by the following procedures:  

1)	the method of digital improvement of energy resolution without the reduction of the efficiency of the cascade recording \cite{Sukh};

2)	the algorithm for determination of a sequence of the resolved cascade quanta in any given interval of energies of their primary transitions, with the use of methods and results of nuclear spectroscopy \cite{Popov}.

For the first time, the procedure of extracting the level density and the partial widths of $\gamma$--emission from the $(n,2\gamma)$--reaction investigation was realized in Dubna, Ref.\cite{Bonev} and \cite{Vasi}. From the measured spectrum of the two-step cascade, the intensity of $I_{\gamma\gamma}(E_1)$ is determined, which links the neutron resonance $\lambda$ (with the excitation energy $E_{ex}$) to the group of final levels $f$ via intermediate levels $i$, by the dipole transitions. This can be represented by the following equation:
\begin{equation}
I_{\gamma\gamma}(E_1)=\sum_{\lambda,f}\sum_{i} \frac{\Gamma_{\lambda i}}{\Gamma_{\lambda}} \frac{\Gamma_{i f}}{\Gamma_{i}} = \sum_{\lambda,f}\sum_j\frac{\Gamma_{\lambda j}}{\left\langle \Gamma_{\lambda j} \right \rangle m_{\lambda j}}n_{j} \frac{\Gamma_{j f}}{\left\langle \Gamma_{j f} \right \rangle m_{j f}}
\label{1}
\end{equation}                                                                                       
where the sum of partial widths of primary transitions $\Sigma_i\Gamma_{\lambda i}$ to $M_{\lambda i}$ intermediate levels i  is $\left\langle \Gamma_{\lambda i} \right\rangle M_{\lambda i}$, and this sum for secondary transitions to $m_{if}$  intermediate levels is $\left\langle \Gamma_{if} \right\rangle m_{if}$ (as $\left\langle \Gamma_{\lambda i} \right\rangle = \sum_i \Gamma_{\lambda i}/M_{\lambda i}$ and $\left\langle \Gamma_{if}\right\rangle = \sum_i \Gamma_{if}/m_{if}$). The sums of intermediate levels in small energy intervals $\Delta E_j$  are $n_j = \rho \Delta E_j$. The branching ratios for primary $\left[ \Gamma_{\lambda j}/\left(\left\langle \Gamma_{\lambda j}\right\rangle M_j\right)\right]$ and secondary $\left[ \Gamma_{jf}/\left(\left\langle \Gamma_{jf}\right\rangle m_{\lambda j}\right)\right]$ transitions are fixed for each $\Delta E_j$. 
\begin{figure}[] 
\includegraphics[scale=0.2]{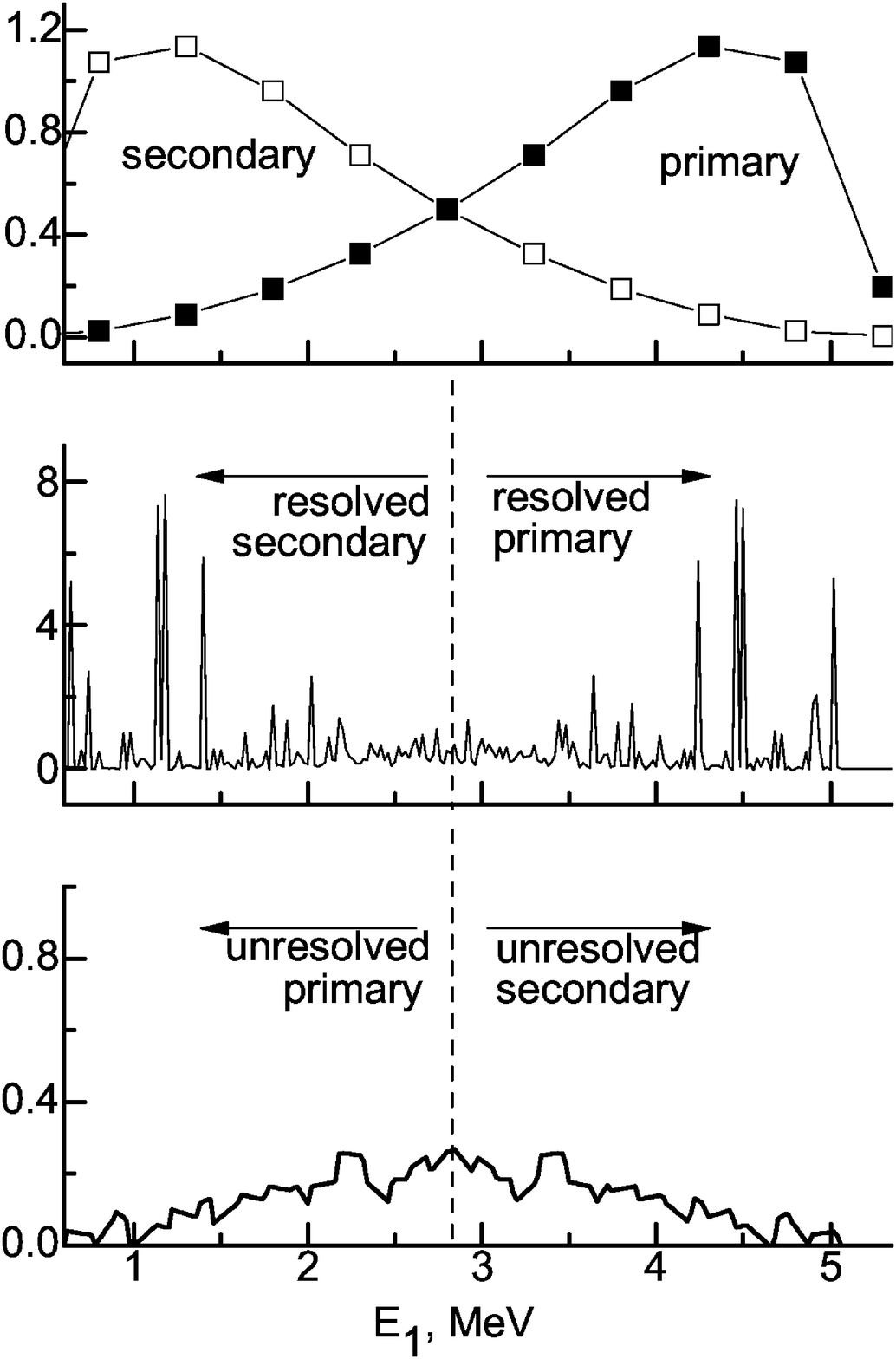}
\centering
\caption{\label{Fig1}
Distribution of the intensity of $5731$ keV cascade for $^{185}W$ calculated by \cite{Kadme,Dilg} models (top picture), with $500$ keV averaging energy interval. Experimental distribution of resolved peaks of the cascade transitions (middle picture) and unresolved continuum (bottom picture), averaging energy interval is $10$ keV.}
\end{figure}
\begin{figure}[h] 
\includegraphics[scale=0.6]{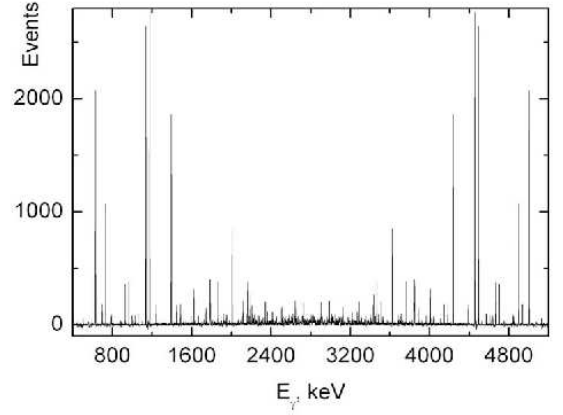}
\centering
\caption{\label{Fig2}
Experimental distribution of intensities of two-step cascades between a neutron resonance and the first excited state of $^{185}W$ taking into account detector efficiency. The spectrum is normalized on the sum of recorded events \cite{Bonda}.}
\end{figure}
\begin{figure}[] 
\includegraphics[scale=0.2]{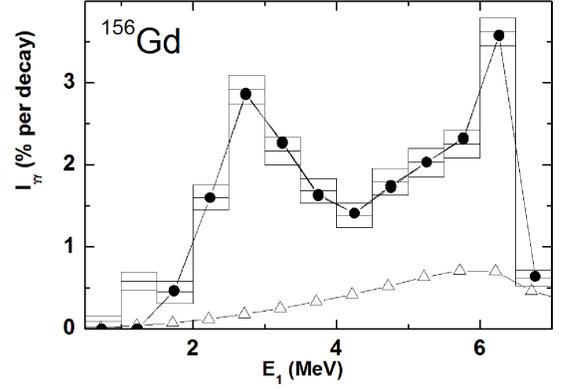}
\centering
\caption{\label{Fig3}
Experimental cascade intensity (histogram) and its uncertainties for $^{156}Gd$ as function of primary cascade quanta $E_1$. Points are the best fit of the presented practical model; triangles are a calculation of $I_{\gamma\gamma}$ using models of Ref. \cite{Kadme,Dilg}. Recorded threshold for cascade gammas is $E_{\gamma} =520$ keV.
}
\end{figure}
\begin{figure}[h] 
\includegraphics[scale=0.67]{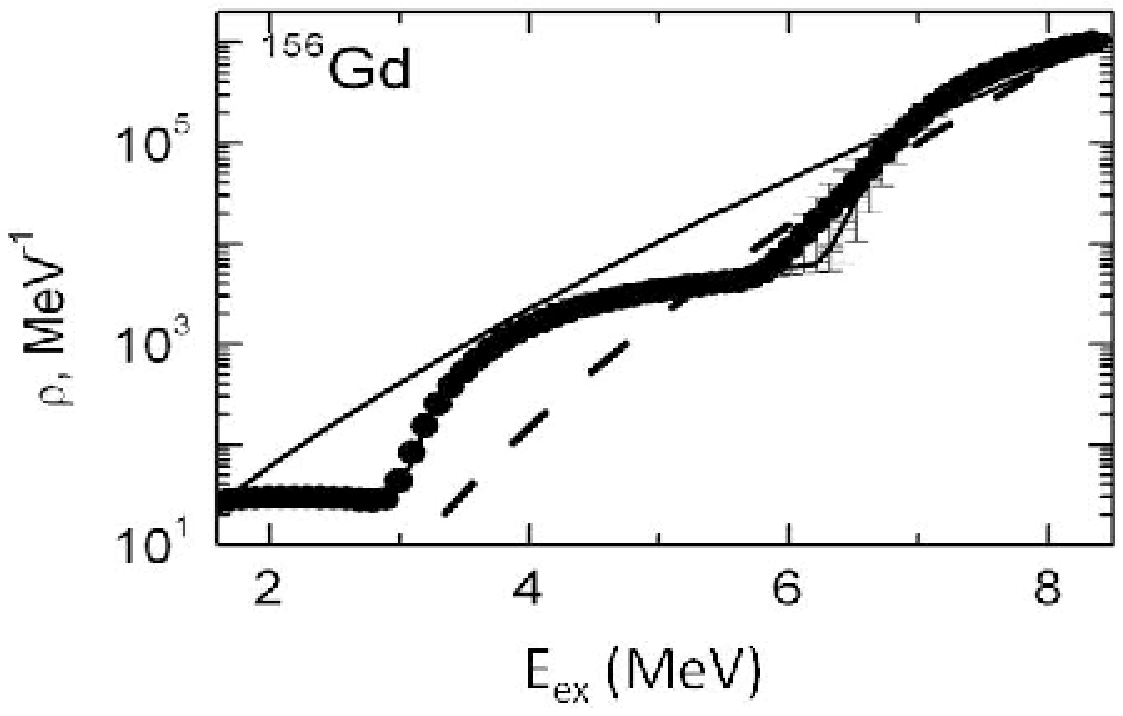}
\includegraphics[scale=0.2]{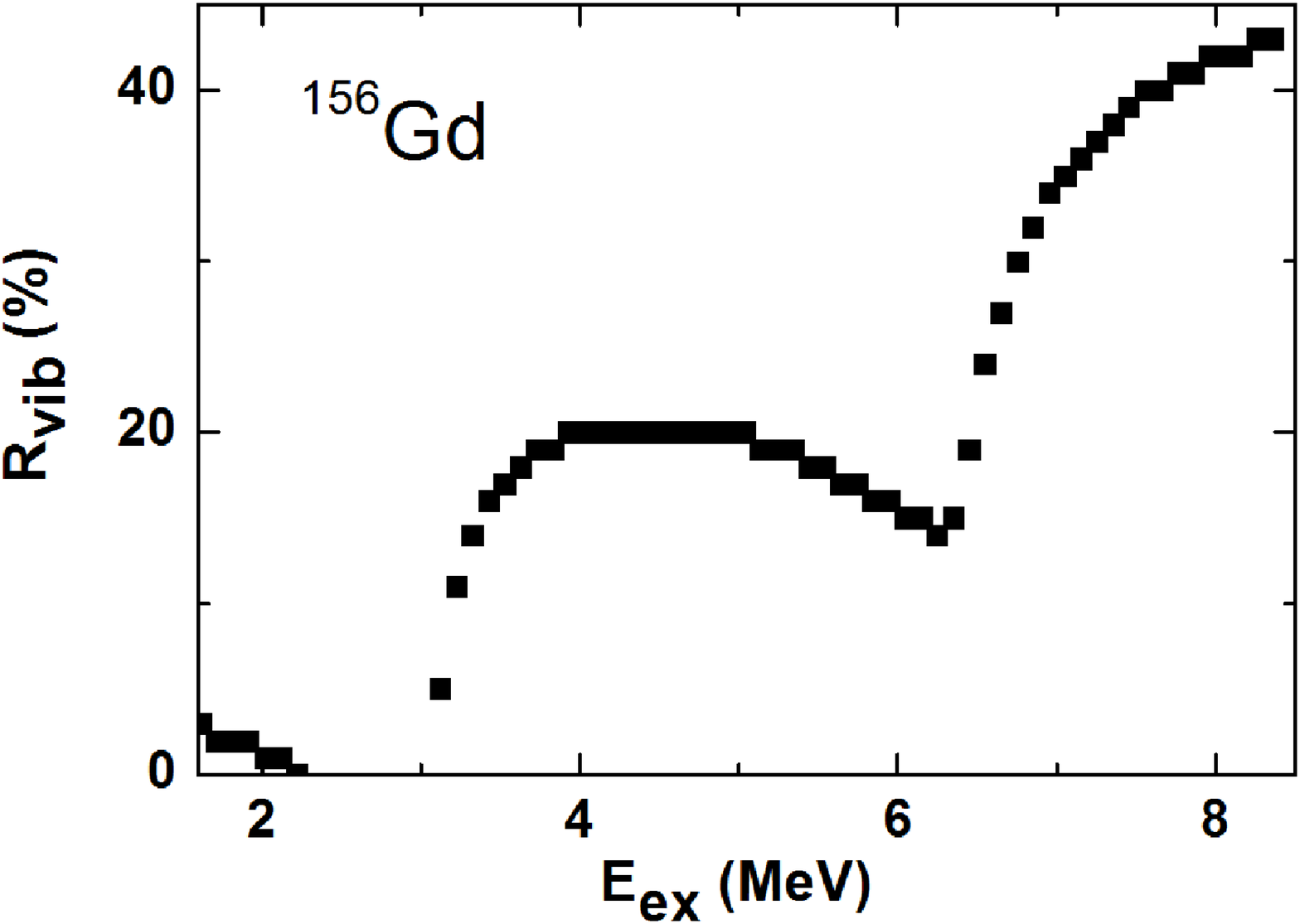}
\centering
\caption{\label{Fig4}
Level density of $^{156}$Gd. In the top picture: points are the best fit of level density (uncertainties -- scatter of fits for different sets of initial parameters); dashed and solid lines are the level density calculated using the model of Ref.  \cite{Dilg}; with taking into account the shell correction $\delta E$ (\ref{6}) and without $\delta E$, correspondingly. Bottom picture: fitted ratio of density of collective levels to the total level density. 
}
\end{figure}
\begin{figure}[h] 
\includegraphics[scale=0.2]{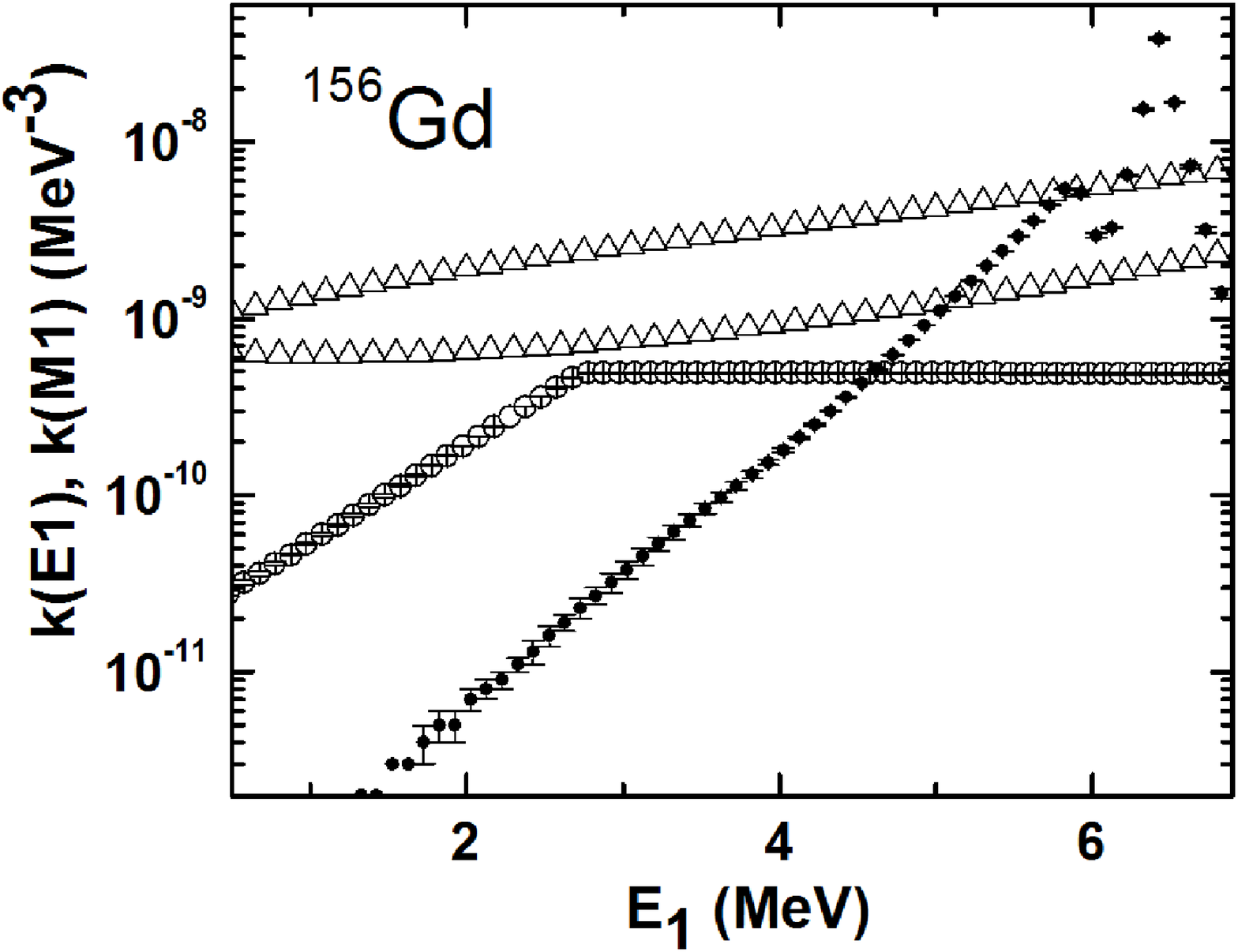}
\includegraphics[scale=0.2]{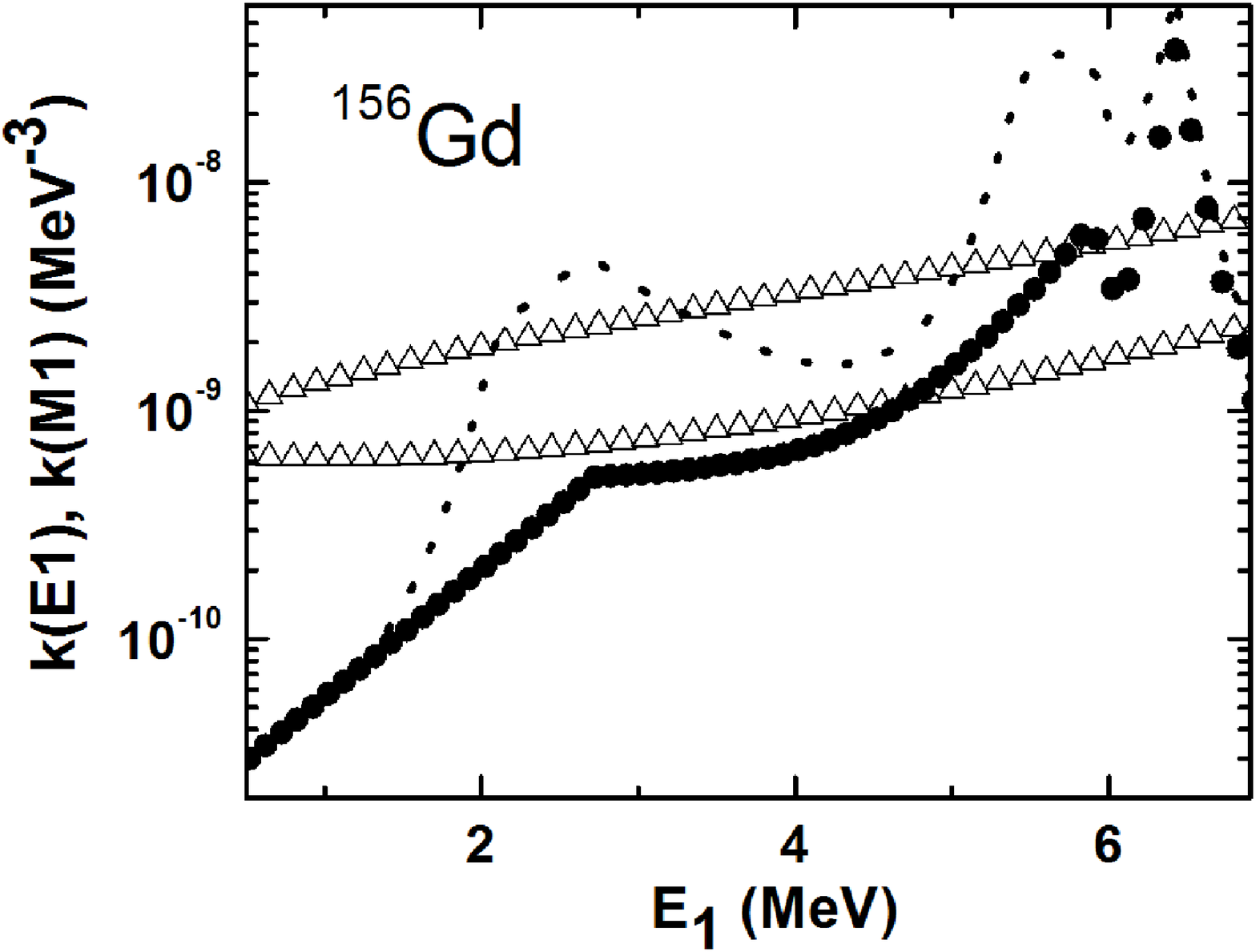}
\centering
\caption{\label{Fig5}
Strength function for $^{156}$Gd. Top panel: solid points are the best fit of the strength function of $E1$ -- transitions; open points are the best fit of the strength function of $M1$ -- transitions. Lower panel: solid points are a sum of $E1$ -- and $M1$ -- strength functions; dash line is the sum of strength functions multiplied by $\rho_{mod}/\rho_{exp}$ ratio (Ref. \cite{Dilg}). Calculations using the model of Ref. \cite{Kadme} (lower triangles) and using the model of Ref. \cite{Axel,Brin} (upper triangles) were fulfilled with $k(M1)= const$.
}
\end{figure}
\begin{figure}[t] 
\includegraphics[scale=0.23]{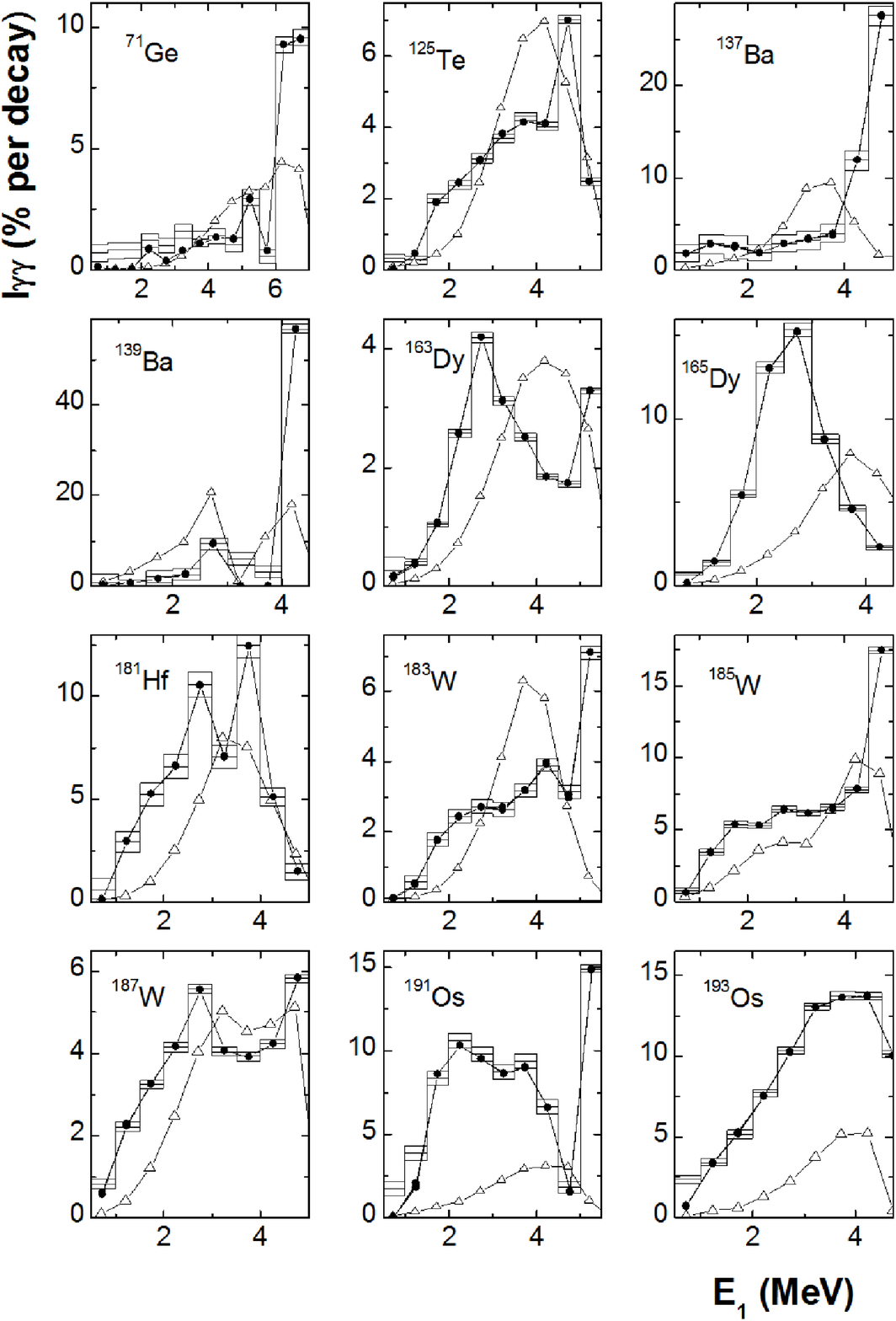}
\centering
\caption{\label{Fig6}
Histograms are the sums of the experimental cascade’s intensities with their small uncertainties in 0.5 MeV bins for even-odd nuclei. Full points are the best fits, triangles are the spectra calculated with the models from Ref. \cite{Kadme,Dilg} with $k(M1) = const$.
}
\end{figure}
\begin{figure}[t] 
\includegraphics[scale=0.24]{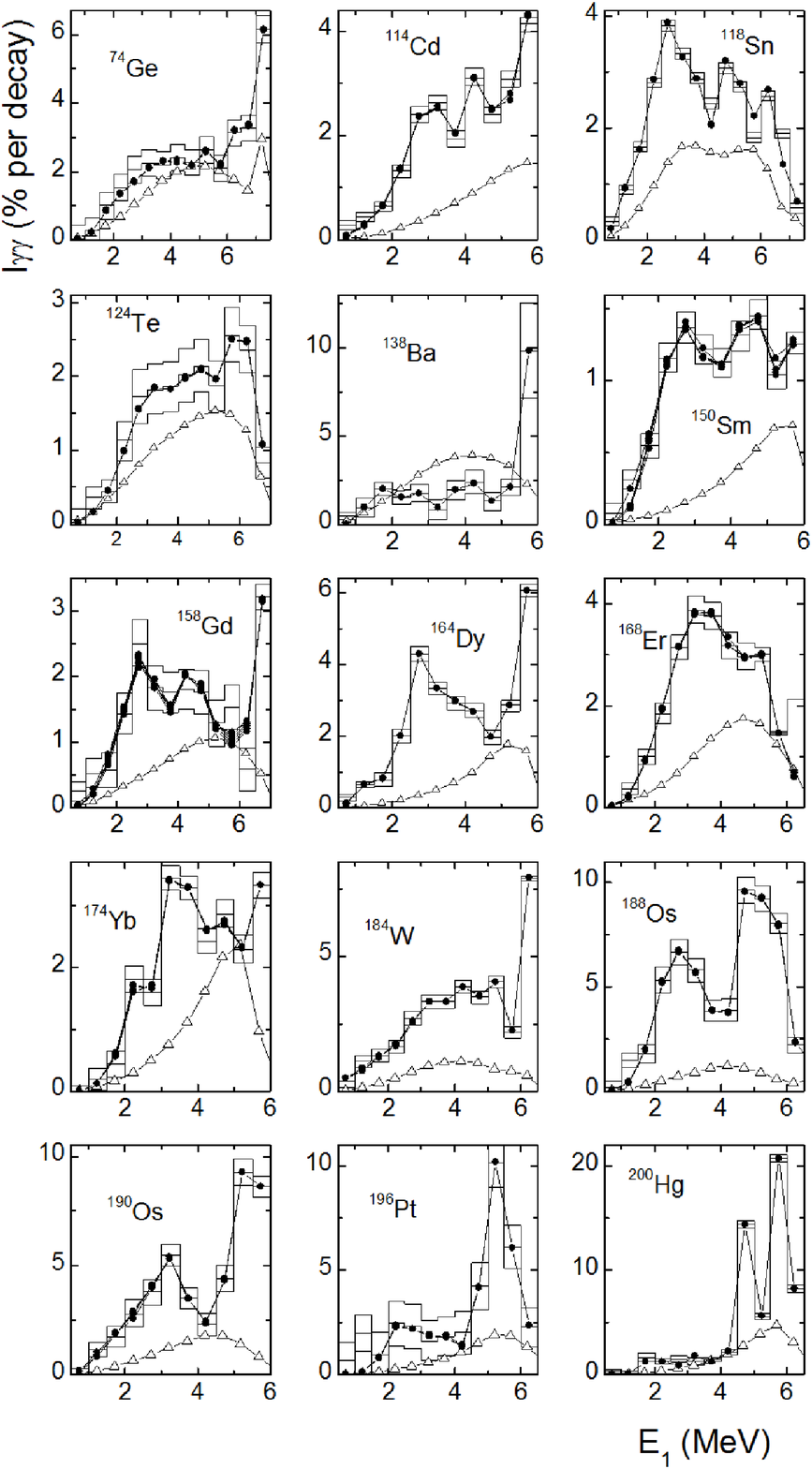}
\centering
\caption{\label{Fig7}
Histograms are the sums of the experimental cascade’s intensities with their small uncertainties in 0.5 MeV bins for even-even nuclei. Full points are the best fits, triangles are the spectra calculated with the models from Ref. \cite{Kadme,Dilg} with $k(M1) = const$.
}
\end{figure}
\begin{figure}[t] 
\includegraphics[scale=0.24]{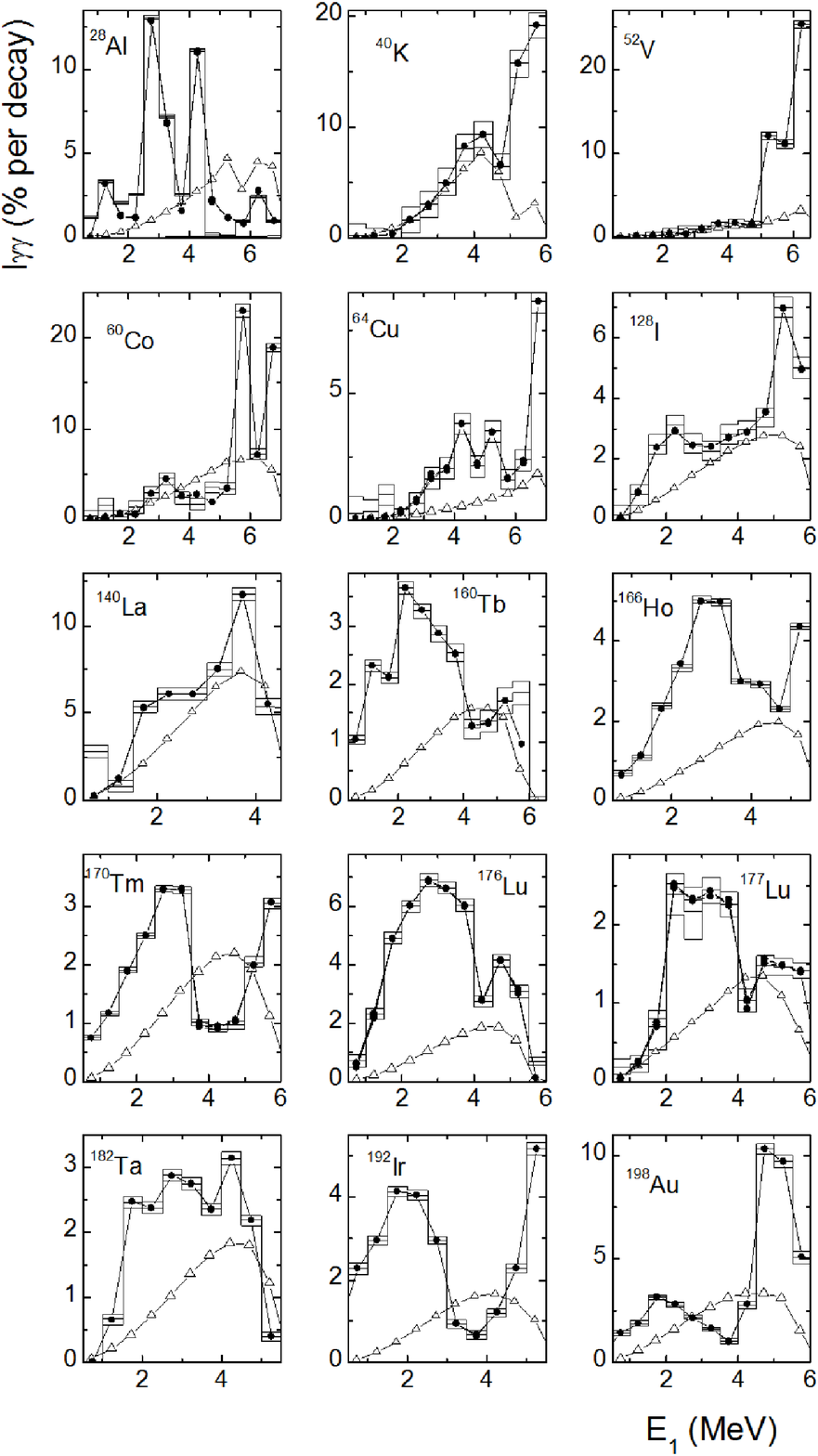}
\centering
\caption{\label{Fig8}
Histograms are the sums of the experimental cascade’s intensities with their small uncertainties in 0.5 MeV bins for odd-odd nuclei. Full points are the best fits, triangles are the spectra calculated with the models from Ref. \cite{Kadme,Dilg} with $k(M1) = const$.
}
\end{figure}
\subsection{The $I_{\gamma\gamma}(E_1)$ spectra preparation}
\label{spectra}
The division of the experimental spectra into two mirror distributions (dependent on the energies of only primary, $E_1$, and only secondary, $E_2$, cascade gamma-quanta) is performed using spectroscopic information. The dividing procedure \cite{Bonev} is based on two facts:

1)	the shapes of the intensity dependencies on energy, for primary and secondary transitions in the same cascade, are mirror --symmetrical;

2)	the resolved peaks of the intensity spectrum of the two-step cascade contain no less than half of the total intensity (this fact was confirmed experimentally for all investigated nuclei).

Fig. \ref{Fig1} illustrates a possibility of such a division of the spectra of intensity. In the top  panel of Fig.1 the calculated intensity distribution of the cascade with the total energy $E_1 + E_2 = 5731$ keV for $^{185}W$ \cite{Bonda} is presented, as an example. For our calculations we used the back shift Fermi-gas model \cite{Dilg} and the  model of \cite{Kadme}. The calculated intensity makes evident division of the spectrum into two parts. As it can be seen, peaks of primary and secondary $\gamma$--transitions of the cascade are located symmetrically, in opposite parts of the interval $E_{\gamma}$ of all $\gamma$--quanta energies (for example see Fig. \ref{Fig2}). 

The experimental distribution of the intensity of resolved peaks and the continuum of unresolved peaks are both shown in Fig. \ref{Fig1} in the middle and in the bottom panel, respectively. A combination of ``primary resolved'' and ``primary unresolved'' parts of the cascade intensity is just a desired $I_{\gamma\gamma}=f(E_1)$ distribution. The choice of the ``unresolved'' spectrum part, where primary transitions are located, was done based on the fact that the total level density increases with the growth of  the excitation energy.

Calculated intensity of this cascade is about $5 \%$ per decay, and the corresponding experimental value is about $11 \%$ (at that, resolved peaks account for $8.4 \%$ of the experimental intensity, and unresolved continuum contains $2.6 \%$ of the intensity). In addition to that, the shape of the experimental distribution noticeably differs from the calculated one, i.e. description of the intensity spectrum by the statistical model is not satisfactory. As in the center of the experimental spectrum (near the half of the neutron binding energy $0.5B_n$) the intensity is noticeably smaller than for the calculated spectrum, so we have a good reason to think that the separation of $I_{\gamma\gamma}(E_1)$ from the experiment spectrum was performed with a higher accuracy than it could be expected in the framework of the statistical model of the nucleus.

The experiment modeling shows that a methodical error of this dividing procedure is caused only by the inaccurate allocation of some cascades (when $E_2>E_1$) near $0.5B_n$, but it does not change the sum of intensities. 
\subsection{Location areas of nuclear parameters}
\label{parameters}
The system of nonlinear equations (\ref{1}) designed for the search of unknown functions $\rho = f(E_{ex})$ and $\Gamma = \Phi(E_1)$ is completely degenerate. Nevertheless, these functions can be defined, but only as possible values in some finite areas. Because of nonlinearity of these functions, their values cannot be infinite. When the procedure for extracting $\rho$ and $\Gamma$ values was created \cite{Vasi}, a set of possible functions $\rho = f (E_{ex})$ and $\Gamma = \Phi(E_1)$, that describe the $I_{\gamma\gamma}$ intensity with practically zero uncertainty, was already specified. At an arbitrary choice of the initial $\rho$ and $\Gamma$ values for fitting the system (\ref{1}) we used, in particular, the model of the Fermi-gas as well as the extrapolation of the tail of the Giant dipole resonance. Small, local distortions of the $\rho$ and $\Gamma$ functions were made in each iteration step in order to get a minimal $\chi^2$. In such a way, this reusable procedure was done with different independent initial $\rho$ and $\Gamma$ values and deviations of the random components of the correction vector until $\chi^2$ minimization was reached. This approach is rather stable, if a noticeable anti--correlation between the $\rho$ and $\Gamma$ values is absent, which is ensured by the branching coefficient for the second step of the cascade, that in turn depends on the ratio of partial width $\Gamma_{if}$ of the secondary transition to the total gamma-width $\Gamma_i$ of the decayed intermediate levels $i$ (see eq. \ref{1}).  Different energy dependencies for the spectra of primary quanta and the secondary ones, with the branching coefficients, allow us to bound the area of random $\rho$ and $\Gamma$ functions. A well-defined step-like structure of the level density \cite{Vasi,Sukh1} has resulted from the fittings with any initial parameters.

Such step-like structure of the level density (it can be explained by breaking of Cooper pairs of nucleons in the nucleus) nowise contradicts the smoothness of the experimental spectra obtained from the nucleon reactions, if the $\rho$ and $\Gamma$ values are connected and their product is a smooth function. Nevertheless, in this case, the location areas of the $\rho$ and $\Gamma$ functions (for an accurate description of the experimental intensity) essentially enlarge what was shown in Ref. \cite{Jovan}.

The relative uncertainties, $\delta \rho/\rho$ and $\delta\Gamma/\Gamma$, always exceed $\delta I_{\gamma\gamma}/I_{\gamma\gamma}$. For the lowest energies of the primary transitions of the cascades, such excess may even reach several orders of magnitude. However, it is obligatory to analyze the real transfer coefficients of the uncertainties of the functions $\rho = f(E_{ex})$ and $\Gamma = \Phi(E_1)$ to the $I_{\gamma\gamma}(E_1)$ uncertainty. When the accuracy of $I_{\gamma\gamma}(E_1)$ description is about a few percents, as it has resulted from our analysis (see Figs. \ref{Fig3}, \ref{Fig5}, \ref{Fig6}, \ref{Fig7}, \ref{Fig8}), the accuracy of the $\rho$ and $\Gamma$ determination will be a few tens of percent. 
\section{Basis of the proposed model}
\label{Basis of the proposed model}
The development and refinement of the model presented in \cite{Vasi} was done at the Frank Laboratory of Neutron Physics (FLNP), JINR \cite{Sukho2,Mitsy}. 
\subsection{The level density}
\label{Model of level density}
An expression for the density $\rho_l$ for the levels of fermionic type was taken from the model of density $\Omega_n$ of $n$-quasi-particle states \cite{Strut}:
\begin{equation}
\label{2}
\begin{array}{ll}
\rho_l=\frac{\left(2J+1\right)\cdot \exp\left(-\left(J+1/2\right)^2/2\sigma^2\right)}{2\sqrt{2\pi}\sigma^3}\cdot\Omega_n\left(E_{\mathsf{ex}}\right) \nonumber \\
\\ 
\Omega_n\left(E_{\mathsf{ex}}\right)=\frac{g^n\left(E_{\mathsf{ex}}-U_l\right)^{n-1}}{\left(\left(\frac{n}{2}\right)!\right)^2\left(n-1\right)!}
\end{array}
\end{equation} 
here $E_{\mathsf{ex}}$ is an excitation energy, $g=6a/\pi^2$ is the density of the single-particle states near the Fermi-surface (the value $a$ is taken from the back-shifted Fermi-gas model \cite{Strut,Refer}), $U_{l}$ is the energy of the $l$-th Cooper pair breaking threshold. The cut-off factor $\sigma$ of the spin $J$ for the excited state of compound-nucleus above the maximal excitation energy $E_{\mathsf{d}}$ of the "discrete" level area is also taken from the Fermi-gas model.

For a given excitation energy, $E_{\mathsf{ex}}$, the phenomenological coefficient $C_{\mathsf{col}}$ of the collective enhancement of the vibrational level density (or both vibrational and rotational ones for deformed nuclei) is determined relaying on a theoretical description that can be found in  Ref. \cite{Igna}. This description gives the following equation:
\begin{equation}
\label{3}
C_{\mathsf{col}}=A_{\mathsf{l}}\cdot \exp\left(\sqrt{\left(E_{\mathsf{ex}}-U_l\right)/E_{\nu}}-\left(E_{\mathsf{ex}}-U_l\right)/E_{\mu} \right)+\beta
\end{equation} 
where $A_{l}$ are parameters of densities of the vibrational levels above the breaking point of each $l$-th Cooper pair, $E_{\mu}$ and $E_{\nu}$ determine a change in the nuclear entropy and a change of the quasi-particles excitation energies, correspondingly. The coefficients $A_{l}$ for different pairs are fitted independently, as performed in \cite{Sukho2,Mitsy,Sukho3}. Coefficient $\beta$ is used at a description of the rotation level density. 
\subsection{The radiative strength function}
\label{Model of radiative strength function}
Radiative strength functions: 
\begin{equation}
\label{4}
k=\Gamma/\left(A^{2/3}E_\gamma^3D_\lambda\right)
\end{equation} 
for $E1$--transitions are determined in Ref. \cite{Kadme}:
\begin{equation}
\label{5}
\begin{array}{l l}
k\left(E1,E_\gamma\right)=w\frac{1}{3\pi^2\hbar^2c^2A^{2/3}}\frac{\sigma_G\Gamma^2_G\left(E^2_\gamma+\kappa4\pi^2T^2\right)}{\left(E^2_\gamma+E^2_G\right)^2+E_\gamma^2\Gamma^2_\gamma}+ \\
\\
+P\delta^-exp\left(\alpha_p\left(E_\gamma-E_p\right)\right)+P\delta^+exp\left(\beta_p\left(E_p-E_\gamma\right)\right)
\end{array}
\end{equation}
with thermodynamic temperature $T$, fitting normalization parameter $w$ and coefficient $\kappa$. Cascades with pure quadrupole transitions were not observed in our experiments. And radiation strength functions for $M1$-transitions are determined for fitting in a similar manner. 

The location of the center of the giant dipole resonance $E_{\mathsf{G}}$, its width $\Gamma_{\mathsf{G}}$ and cross section $\sigma_{\mathsf{G}}$ in the maximum are taken from \cite{Diet} for each nucleus. The necessity to add one or several peaks to the strength function is based on the data of Ref. \cite{Sukh4}. The shape of each of the peaks we described in different ways (in presented analysis it was done by two exponential functions). The second summand of Eq. (\ref{5}) corresponds to the left slope of the peak (energies below the maximum), and the third summand is the right slope (energies above the maximum). Position $E_{\mathsf{p}}$ in the energy scale, amplitude $P$ and slope parameters $\alpha_{\mathsf{p}}$ and $\beta_{\mathsf{p}}$ are fitted for each peak independently. 

At $E_{\mathsf{1}}\approx B_{\mathsf{n}}$ the fitted ratios $\Gamma_{M1}/\Gamma_{E1}$ of $E1$- and $M1$-strength functions are normalized to known experimental values, and their sum $\Gamma_\lambda$ is normalized to the total radiation width of the resonance. 
\subsection{Parameters for fitting}
\label{Fitted parameters}
The set of common parameters for fitting (see equations \ref{2}, \ref{3} and \ref{5}) of all cascades in our model is the following:

1) the energies $U_{l}$ of breaking thresholds up to $l=4$;

2) the $E_{\mu}$ and $E_{\nu}$ parameters;
 
3) the independent parameters $A_{l}$ of the density of vibrational levels above the break up threshold $U_{l}$;

4) the coefficients $w$, $\kappa$, $\beta$, $\beta$, $P$, $E_p$, $ \alpha_p$ and $\beta_p$;

5) the ratio $r$ of the levels with negative parity to the total level density.

This set of parameters were used for the description of the intensity $I_{\gamma\gamma}(E_1)$ for 43 nuclei in the mass interval $28 \leq  A \leq 200$, in the framework of the proposed model.     
\begin{figure}[t] 
\includegraphics[scale=0.23]{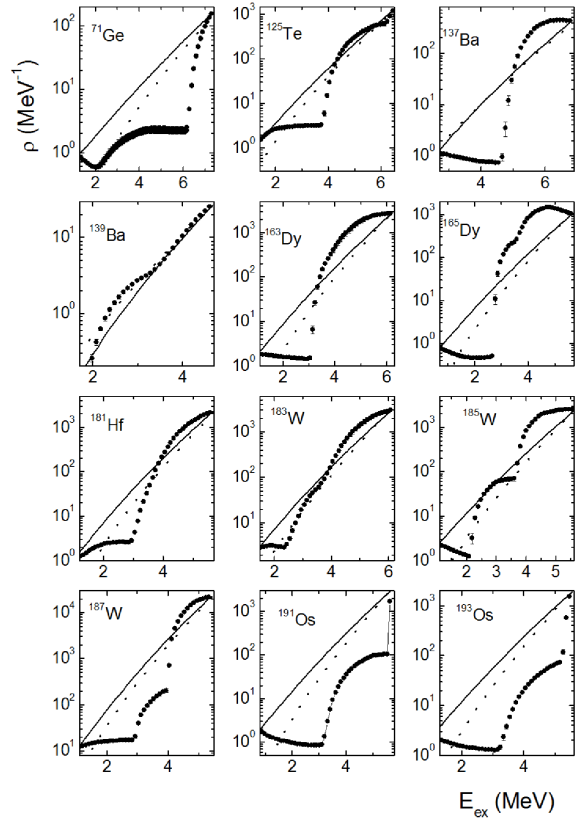}
\centering
\caption{\label{Fig9}
The most probable mean densities of intermediate levels of two-step cascades (full points) for even-odd nuclei and their variations in different fittings with the lowest $\chi^2$. Dashed and solid lines are the level density calculated using the model of Ref. \cite{Dilg}, with taking into account the shell correction $\delta E$ (\ref{6}) and without $\delta E$, correspondingly. 
}
\end{figure}
\begin{figure}[t] 
\includegraphics[scale=0.23]{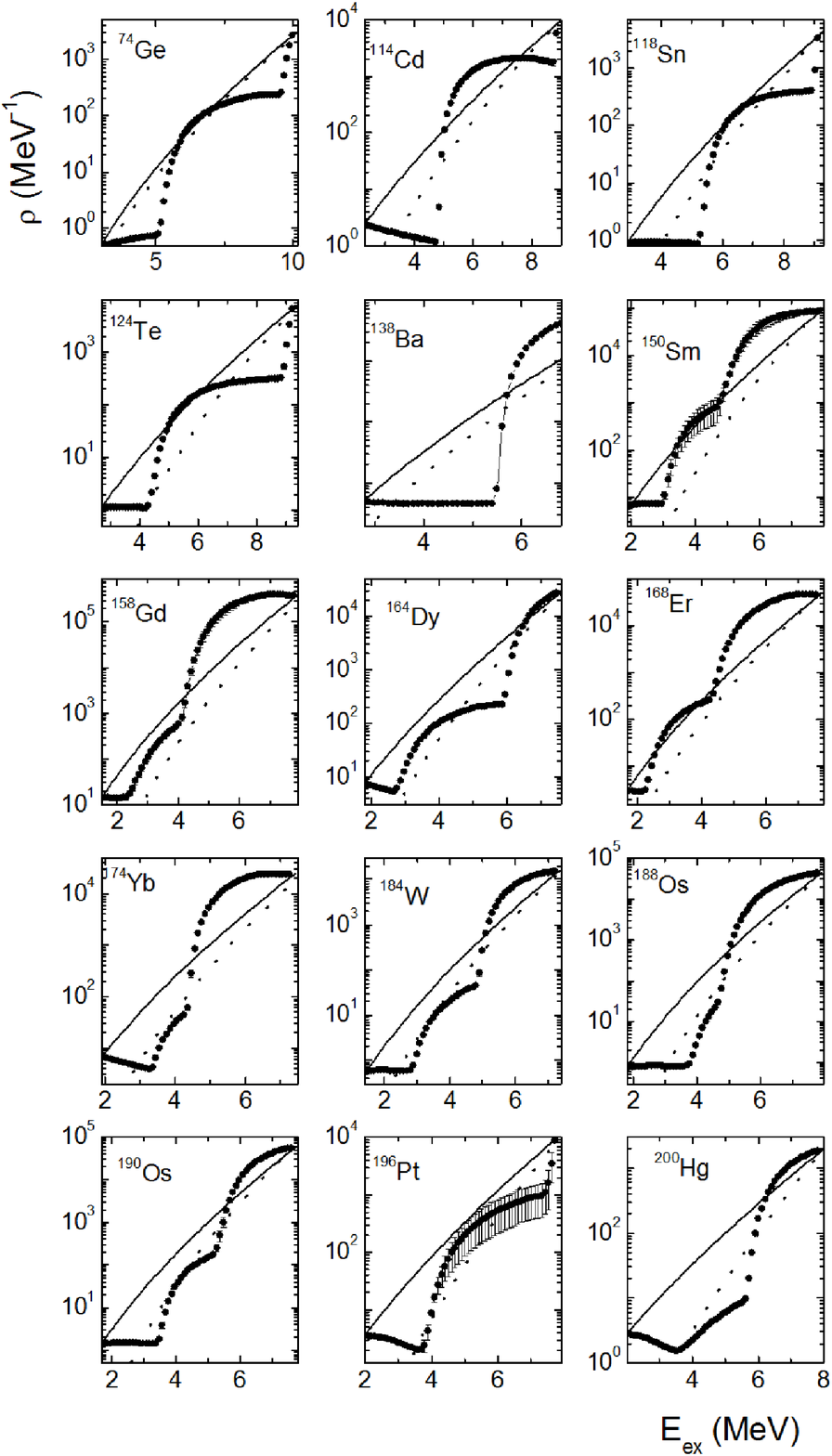}
\centering
\caption{\label{Fig10}
The most probable mean densities of intermediate levels of two-step cascades (full points) for even-even nuclei and their variations in different fittings with the lowest $\chi^2$. Dashed and solid lines are the level density calculated using the model of Ref. \cite{Dilg}, with taking into account the shell correction $\delta E$ (\ref{6}) and without $\delta E$, correspondingly.
}
\end{figure}
\begin{figure}[t] 
\includegraphics[scale=0.23]{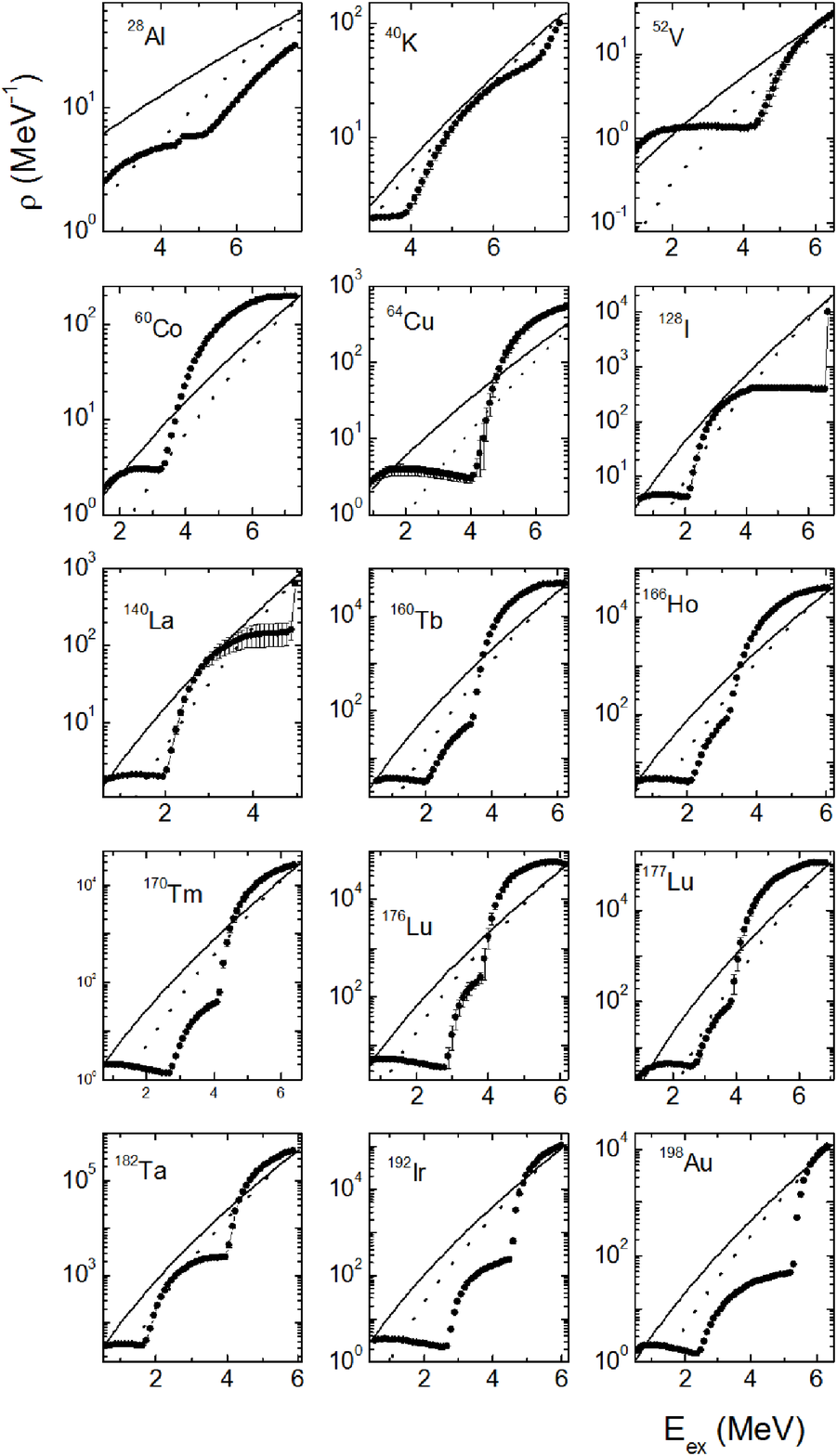}
\centering
\caption{\label{Fig11}
The most probable mean densities of intermediate levels of two-step cascades (full points) for odd-odd nuclei and their variations in different fittings with the lowest $\chi^2$. Dashed and solid lines are the level density calculated using the model of Ref. \cite{Dilg}, with taking into account the shell correction $\delta E$ (\ref{6}) and without $\delta E$, correspondingly.
}
\end{figure}
\begin{figure}[t] 
\includegraphics[scale=0.23]{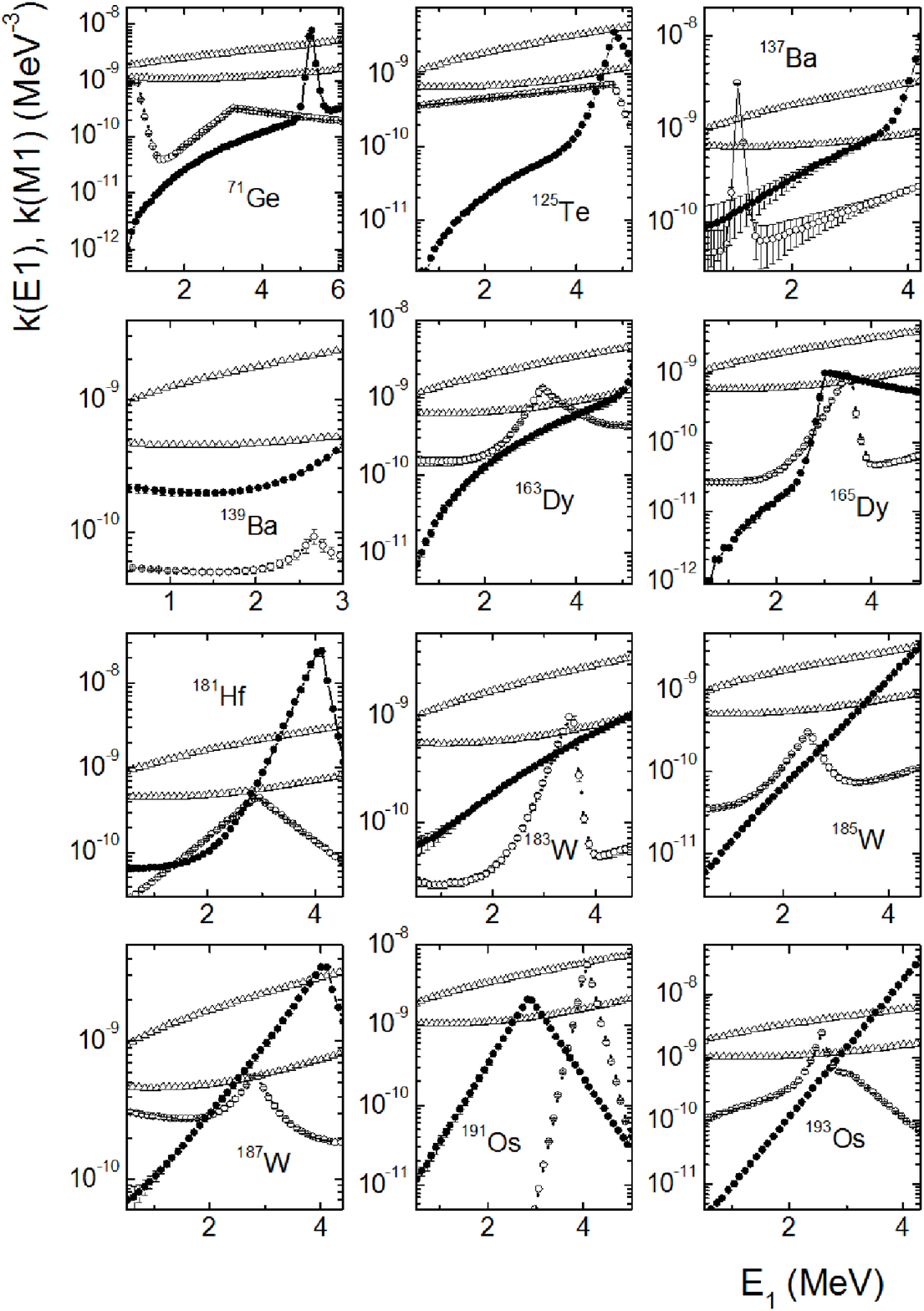}
\centering
\caption{\label{Fig12}
The strength functions of $E1$--transitions (close points) and of $M1$--transitions (open points) for even-odd nuclei (the best fits). The top line of triangles depicts the model calculation from Ref. \cite{Axel}, the bottom line represents the model calculation of Ref. \cite{Kadme} in sum with $k(M1) = const$. 
}
\end{figure}
\begin{figure}[t] 
\includegraphics[scale=0.23]{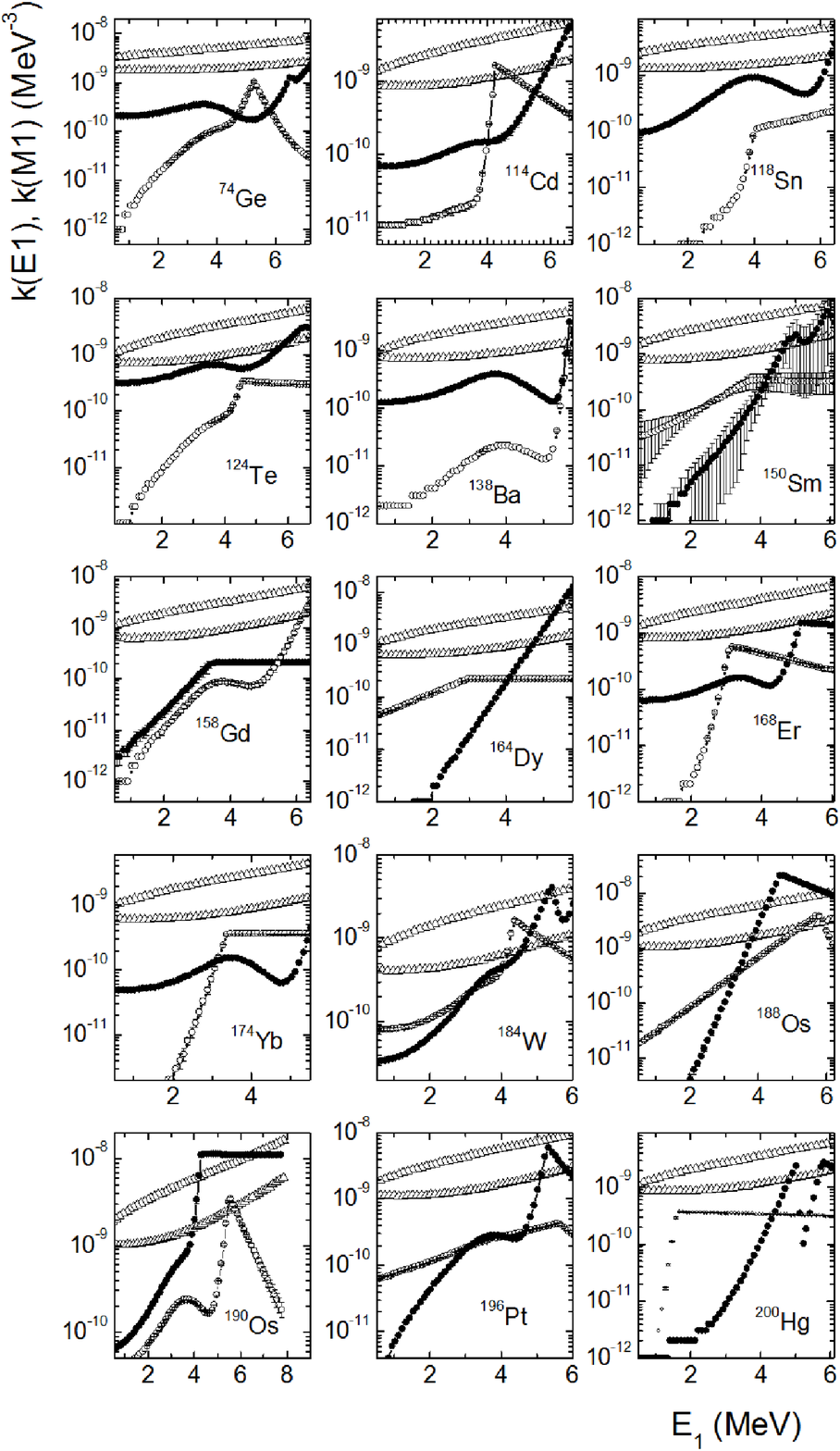}
\centering
\caption{\label{Fig13}
The strength functions of $E1$--transitions (close points) and of $M1$--transitions (open points) for even-even nuclei (the best fits). The top line of triangles depicts the model calculation from Ref. \cite{Axel}, the bottom line represents the model calculation of Ref. \cite{Kadme} in sum with  $k(M1) = const$.
}
\end{figure}
\begin{figure}[t] 
\includegraphics[scale=0.23]{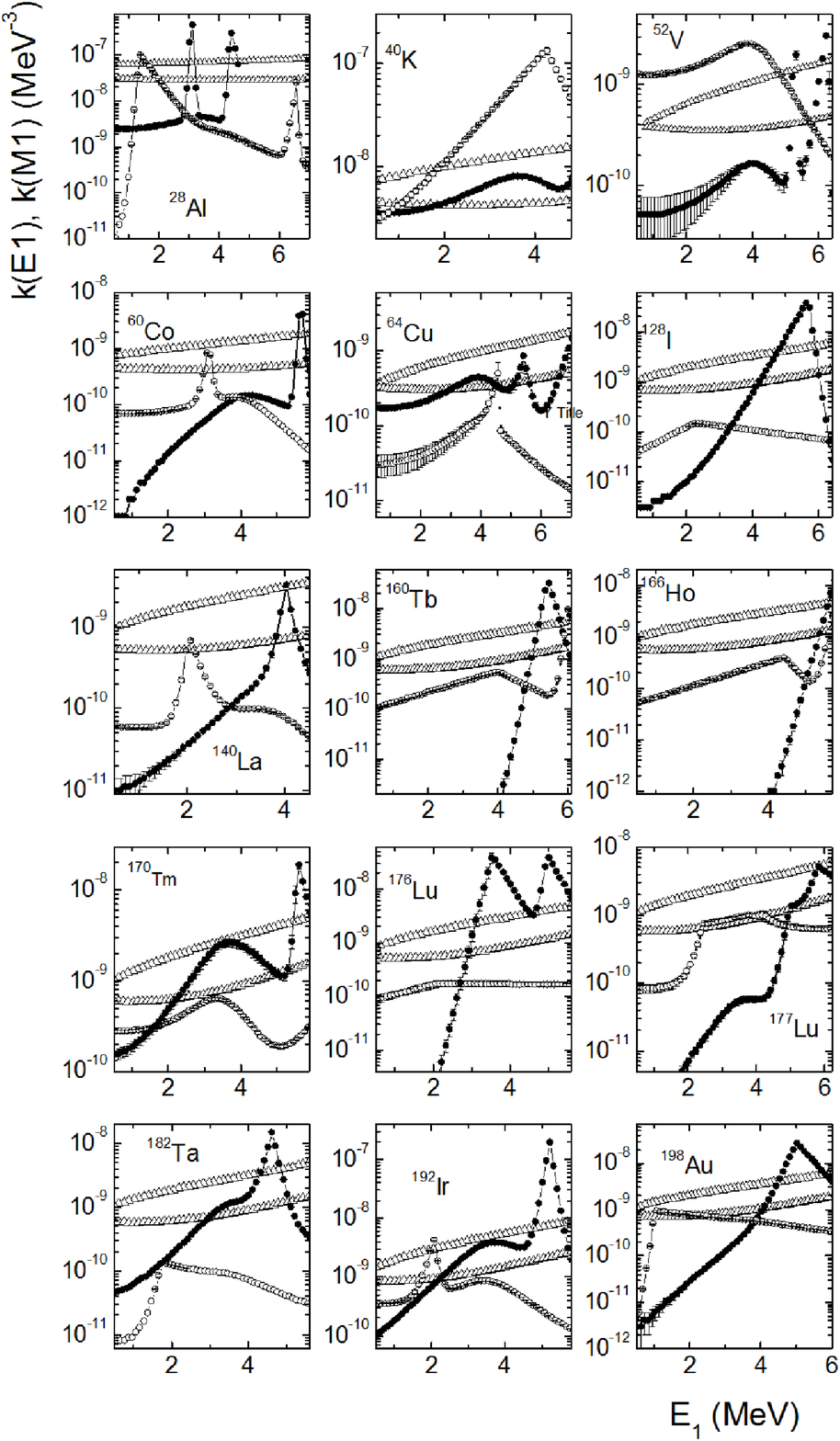}
\centering
\caption{\label{Fig14}
The strength functions of $E1$--transitions (close points) and of $M1$--transitions (open points) for odd-odd nuclei (the best fits). The top line of triangles depicts the model calculation from Ref. \cite{Axel}, the bottom line represents the model calculation of Ref. \cite{Kadme} in sum with  $k(M1) = const$.
}
\end{figure}
\begin{figure}[t] 
\includegraphics[scale=0.23]{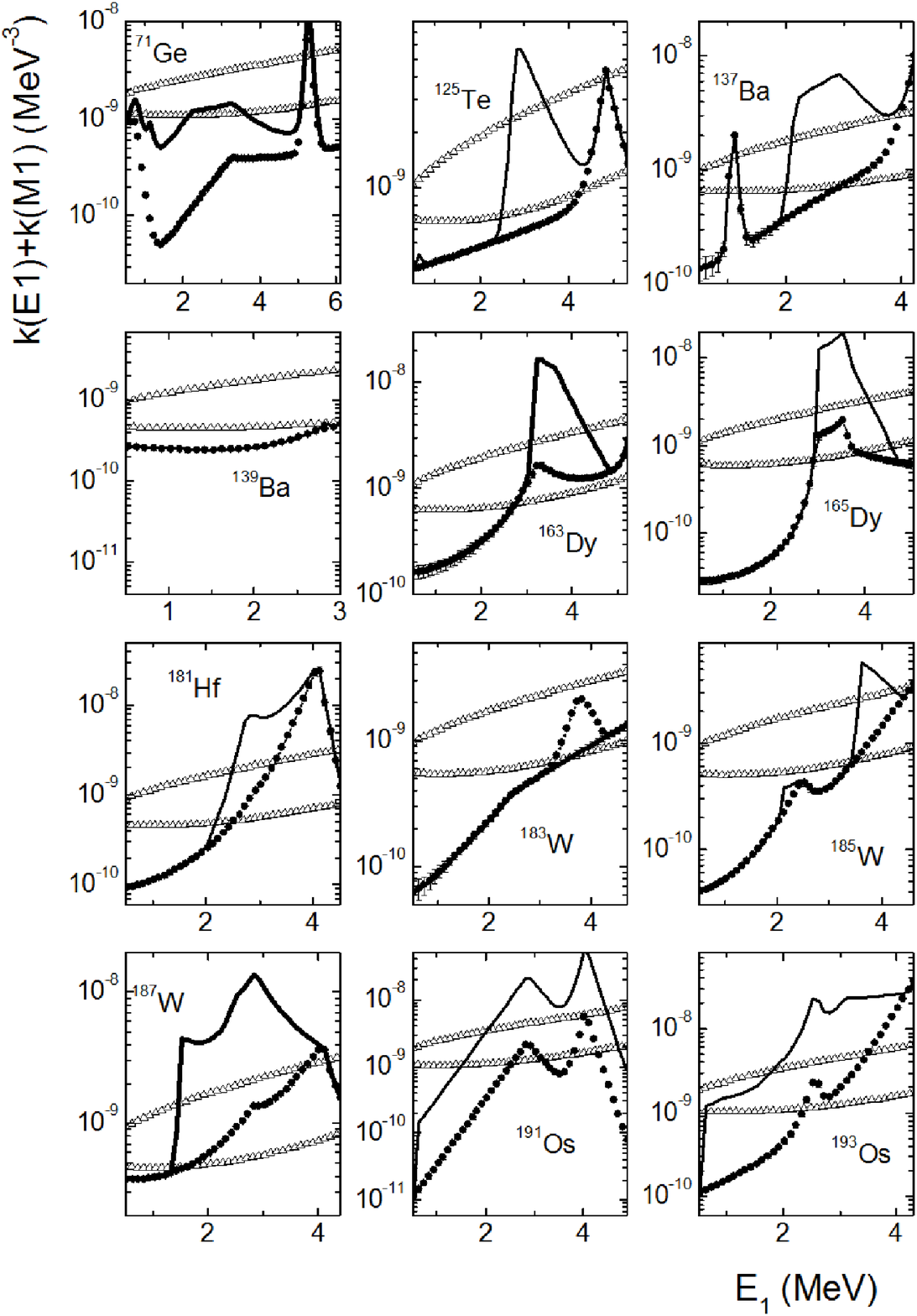}
\centering
\caption{\label{Fig15}
Sum of strength functions of $E1$-- and $M1$--transitions (close points) for even-odd nuclei (the best fits). Solid line is the same multiplied by $\rho_{\mathsf{mod}}/\rho_{\mathsf{exp}}$ ratio (Ref. \cite{Dilg}). The top line of triangles depicts the model calculation from Ref. \cite{Axel}, the bottom line represents the model calculation of Ref. \cite{Kadme} in sum with $k(M1) = const$. 
}
\end{figure}
\begin{figure}[t] 
\includegraphics[scale=0.23]{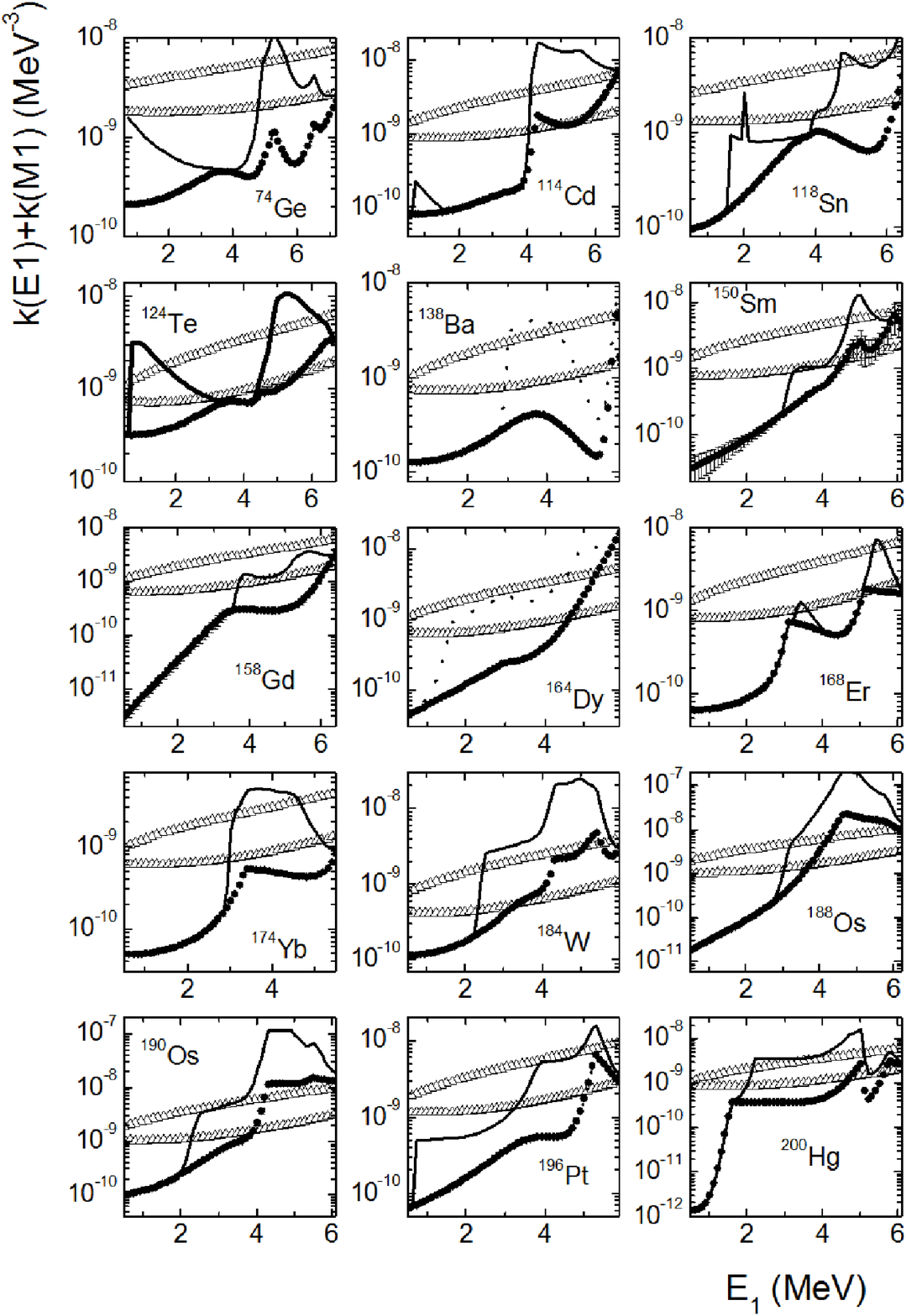}
\centering
\caption{\label{Fig16}
Sum of strength functions of $E1$-- and $M1$--transitions (close points) for even-even nuclei (the best fits). Solid line is the same multiplied by $\rho_{\mathsf{mod}}/\rho_{\mathsf{exp}}$ ratio (Ref. \cite{Dilg}). The top line of triangles depicts the model calculation from Ref. \cite{Axel}, the bottom line represents the model calculation of Ref. \cite{Kadme} in sum with $k(M1) = const$. 
}
\end{figure}
\begin{figure}[t] 
\includegraphics[scale=0.23]{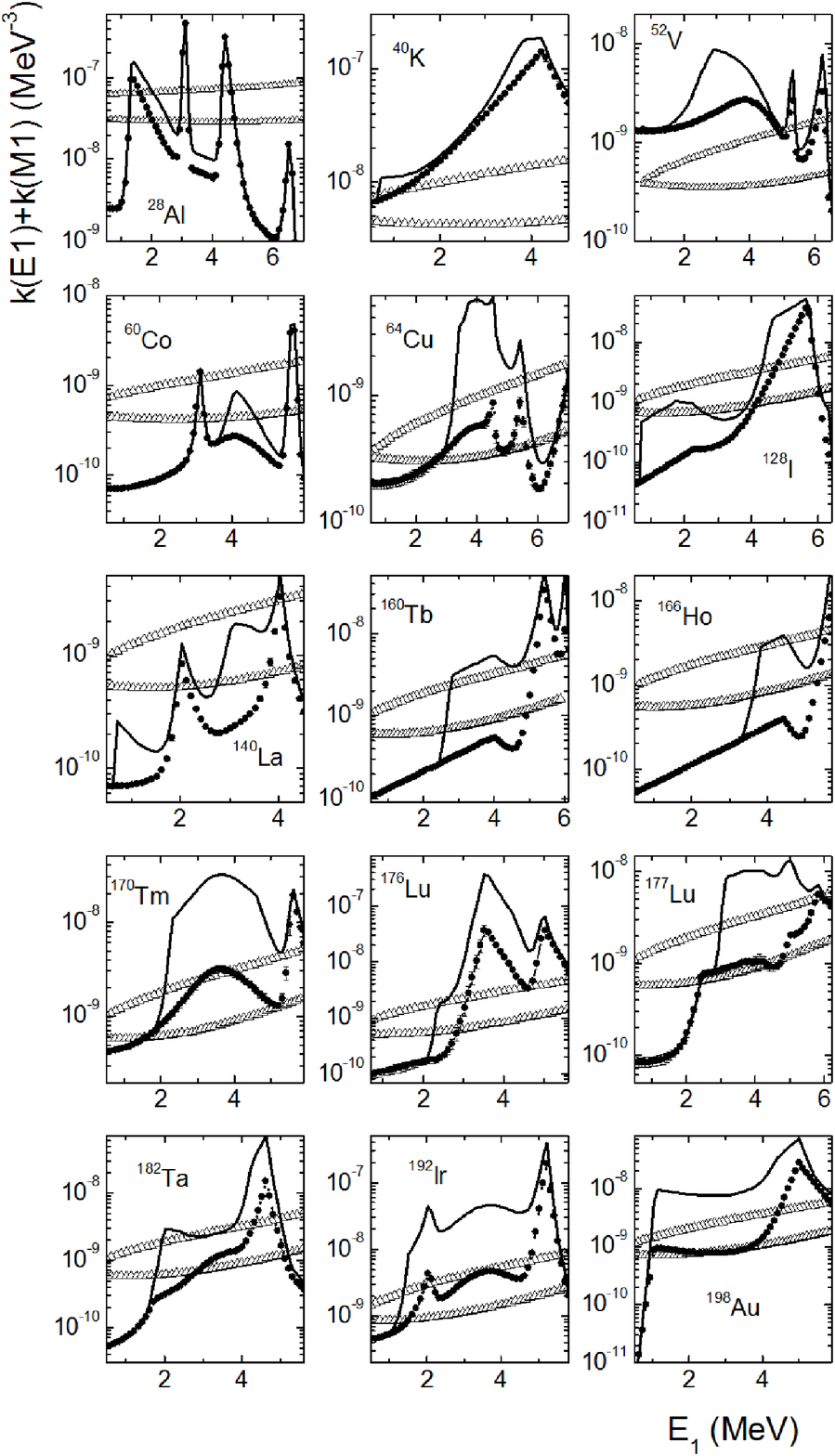}
\centering
\caption{\label{Fig17}
Sum of strength functions of $E1$-- and $M1$--transitions (close points) for odd-odd nuclei (the best fits). Solid line is the same multiplied by $\rho_{\mathsf{mod}}/\rho_{\mathsf{exp}}$ ratio (Ref. \cite{Dilg}). The top line of triangles depicts the model calculation from Ref. \cite{Axel}, the bottom line represents the model calculation of Ref. \cite{Kadme} in sum with $k(M1) = const$. 
}
\end{figure}
\begin{figure}[t] 
\includegraphics[scale=0.23]{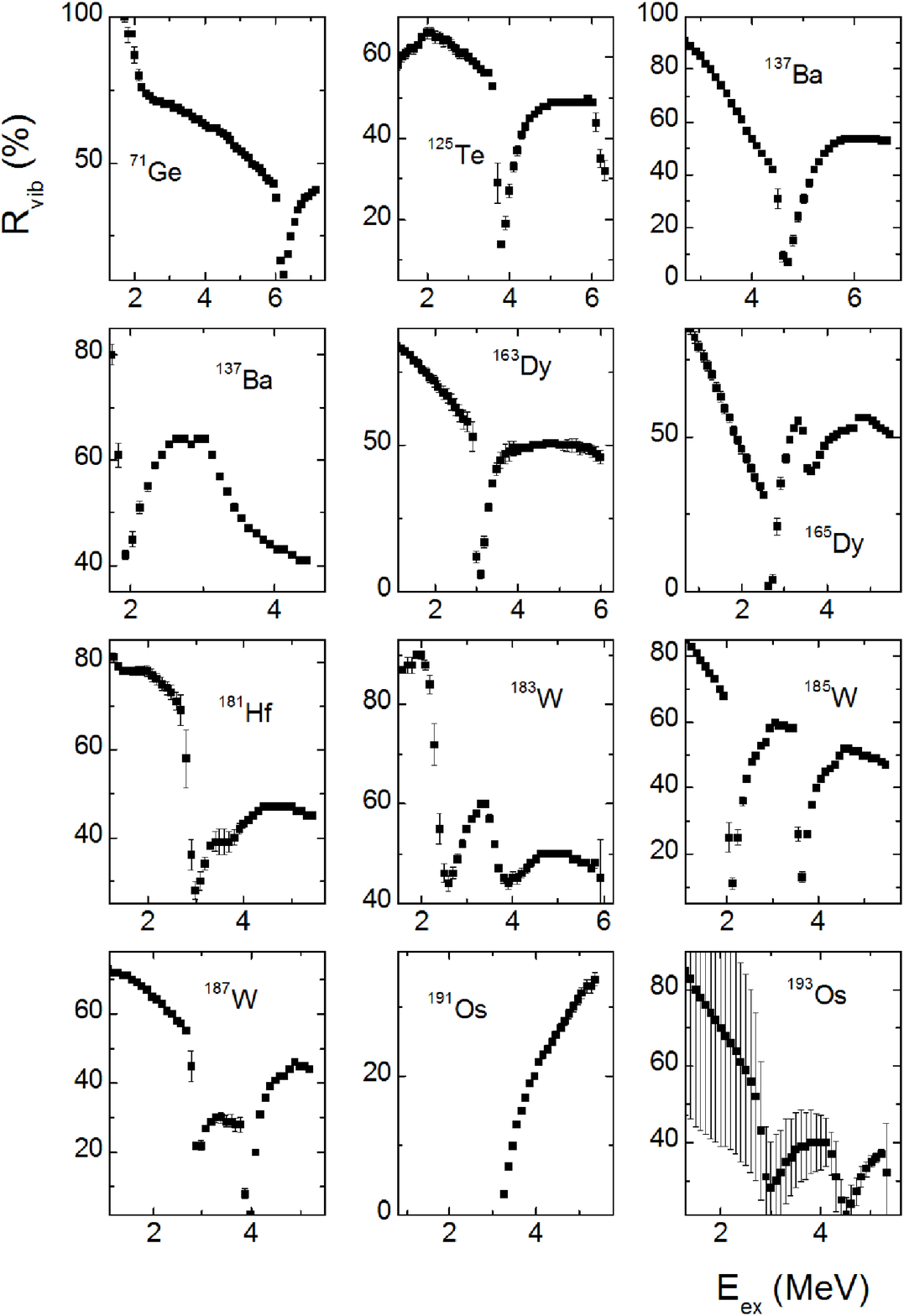}
\centering
\caption{\label{Fig18}
The part of vibrational levels $R_{\mathsf{vib}}$ in the total density of excited levels for even-odd nuclei at excitation energy $E_{\mathsf{ex}}$. 
}
\end{figure}
\begin{figure}[t] 
\includegraphics[scale=0.23]{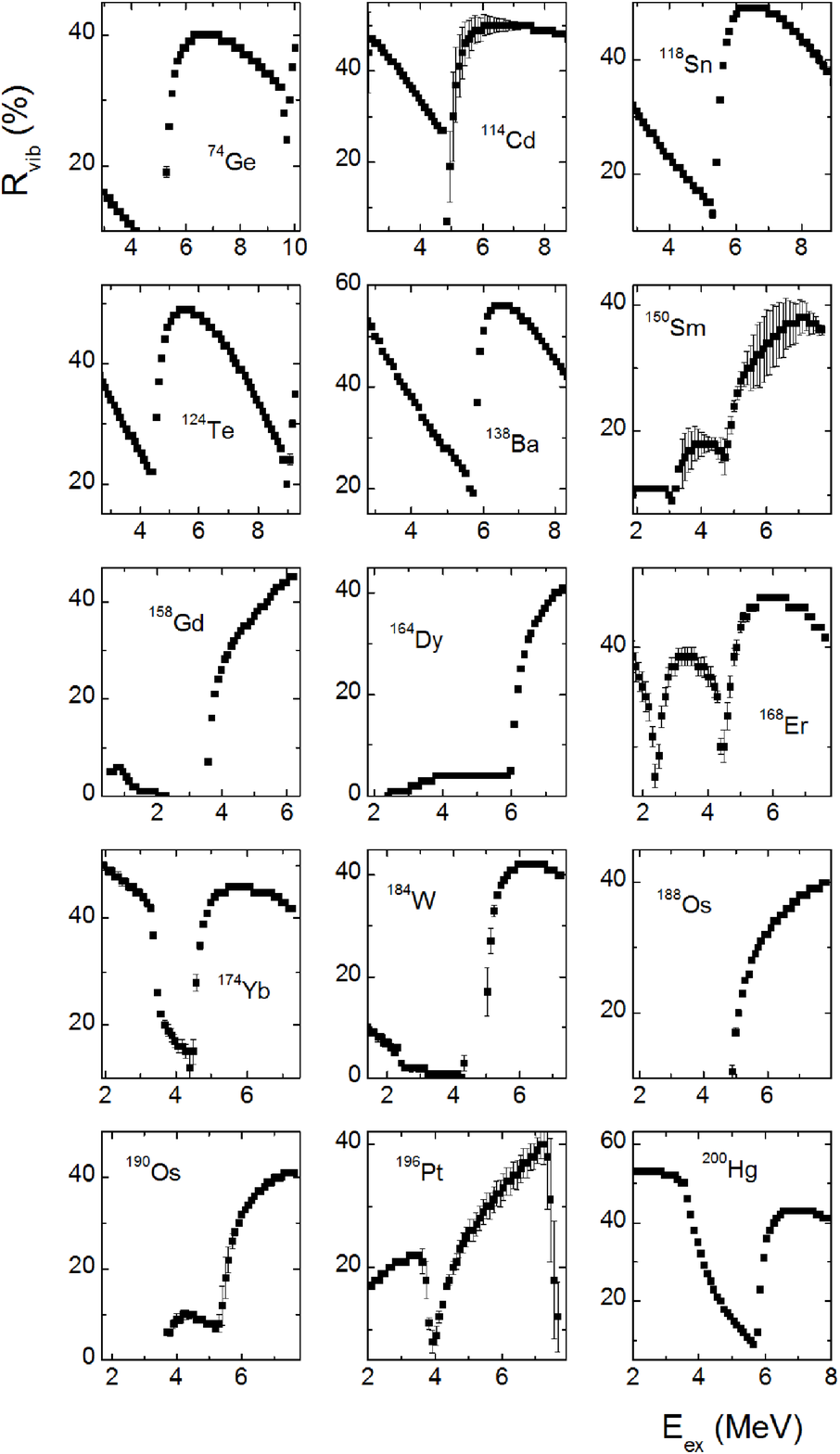}
\centering
\caption{\label{Fig19}
The part of vibrational levels $R_{\mathsf{vib}}$ in the total density of excited levels for even-even nuclei at excitation energy $E_{\mathsf{ex}}$. 
}
\end{figure}
\begin{figure}[t] 
\includegraphics[scale=0.23]{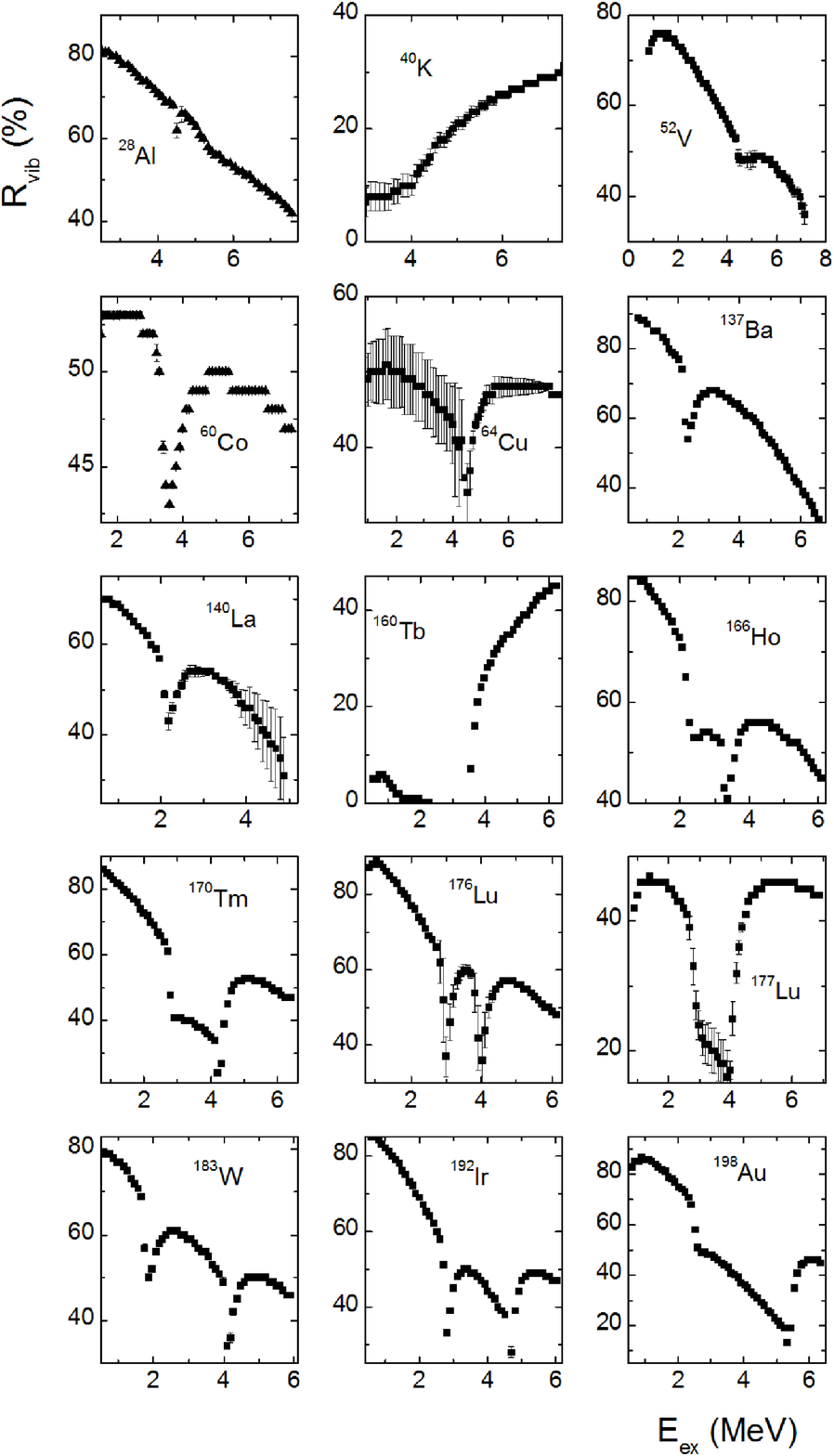}
\centering
\caption{\label{Fig20}
The part of vibrational levels $R_{\mathsf{vib}}$  in the total density of excited levels for odd-odd nuclei at excitation energy $E_{\mathsf{ex}}$. 
}
\end{figure}
%
\section{Practical model improvements}
\label{Practical model improvements}
There is a significant disagreement between the measured cascade intensities and ones calculated by the statistical model \cite{Sukho2,Mitsy}. To obtain the most reliable information about the nuclear matter properties it is necessary to combine several models for $\rho$ and $\Gamma$ \cite{Refer}.   
 
Dubna model is based on the conclusion of the theoretical analysis (Ref. \cite{Malov}) concerning the fragmentation of different quasi-particles states in a nuclear potential, that Cooper pair breaking during nuclear excitation is a sequential process. At that, the Dubna model allows us to examine two opposite hypotheses (the particular cases of above mentioned theory): that the nucleus is a pure fermion system or that there is a phase transition at some excitation energy to the nucleus consisted of bosons.

There are no known fully precise and correct models about the behavior of the nuclear matter in excited nuclei. The singular verifiable hypothesis, realized  when studying nuclear superfluidity, is the increase of the total level density, which grows in such manner, as it is taken into account by $C_{col}$ coefficient. 

At first, in the our practical model, \cite{Sukho2, Mitsy} we assumed that the $E_{\mu}$ and $E_{\nu}$ parameters of the vibrational level density vary for different broken Cooper pairs independently and the density $g$ of the single-particle states is a constant near the Fermi-surface for any given nucleus. However, results from Ref. \cite{Sukho3} showed that $E_{\mu}$ and $E_{\nu}$ can be replaced by the same parameter (i.e. $E_{\mu}$ = $E_{\nu}$), what allowed us to decrease the number of model parameters. Moreover, an analysis of scores of fittings shows that this common parameter can be taken for all of the broken pairs in a given nucleus. 

According to the results of theoretical investigations \cite{Igna} it is necessary to take into account an influence of the shell inhomogeneity of a single-particle spectrum on the obtained $\rho$ and $\Gamma$ values.
\section{Correction for shell inhomogeneities }
\label{Correction for shell inhomogeneities }
The theoretical opinions about the influence of the shell correction $\delta E$ on the density of the quasi-particle levels were tested in this work for all 43 nuclei investigated in Dubna. The testing was performed using the $a(A)$ value, which depends on the excitation energy, included linearly in the parameter of the single-particle density $g$ (see Eq. \ref{2}). For a nucleus with mass $A$ and excitation energy $E_{ex}$, $a(A)$ is expressed, according to Ref. \cite{Igna},  as: 
\begin{equation}
\label{6}
a\left(A\right)=\tilde{a}\left(1+\left(\left(1-exp\left(\gamma E_{\mathsf{ex}}\right)\right)\cdot \delta E/E_{\mathsf{ex}}\right)\right)
\end{equation} 
where asymptotic value equals to $\tilde{a}=0.114\cdot A+0.162\cdot A^{2/3}$ and $\gamma=0.054$. In order to keep an average spacing between neutron resonances \cite{Sukho2,Mitsy,Sukho3}, $\delta$E values slightly varied relative to their evaluations \cite{Igna}. The shell corrections, used at fitting the parameters of the Dubna model, are presented in the Table \ref{Tab1}.
\begin{table*}[]
\centering
\caption{Used in the analysis values (the maximal excitation energy $E_{\mathsf{d}}$ of the "discrete" level area, the energy $E_{\mathsf{max}}$ of the final level of the cascade, the shell correction $\delta E$ on the density of quasi-particle levels and intensity $I_{\gamma\gamma}$ of the two-step cascade) for investigated nuclei.}
\label{Tab1}
\begin{tabular}{|l|l|l|l|l|l|}
\hline
\textbf{}
Nucleus	  & $E_{\mathsf{d}}$(MeV) &	$E_{\mathsf{max}}$(MeV) &	Shell correction $\delta E$ (MeV) &	$I (\%)$ & Spins of state $\lambda$ \\ \hline
$^{28}\mathsf{Al}$	&2.486	&0.972  &-11.1 &49(1) &2,3\\ \hline
$^{40}\mathsf{K}$	  &2.985	&1.64	  &-3.1 &67(3)	&1,2\\ \hline
$^{52}\mathsf{V}$	  &0.846	&0.147	&-5.0 &60(2)	&3,4\\ \hline
$^{60}\mathsf{Co}$	&1.515	&1.5	  &-5.9 &71(3)	&3,4\\ \hline
$^{64}\mathsf{Cu}$	&0.926	&0.278	&-3.2 &30(6)	&1,2 \\ \hline
$^{71}\mathsf{Ge}$	&1.298	&0.0	  &-3.5	&32(2)	&1/2\\ \hline
$^{74}\mathsf{Ge}$	&2.963	&2.165	&-3.0	&36(2)	&4,5\\ \hline
$^{114}\mathsf{Cd}$	&2.316	&0.558	&-1.0	&26(1)	&0,1\\ \hline
$^{118}\mathsf{Sn}$	&2.930	&1.230	&-1.8	&31(1)	&0,1\\ \hline
$^{124}\mathsf{Te}$	&2.702	&0.603	&-0.3	&20(2)	&0,1\\ \hline
$^{125}\mathsf{Te}$	&1.319	&0.671	&-2.3	&31(1)	&1/2\\ \hline
$^{128}\mathsf{I}$	&0.434	&0.434	&-1.0	&33(2)	&2,3\\ \hline
$^{137}\mathsf{Ba}$	&2.662	&0.279	&-6.3	&59(4)	&1/2\\ \hline
$^{138}\mathsf{Ba}$	&2.780	&1.436	&-8.2	&26(5)	&1,2\\ \hline
$^{139}\mathsf{Ba}$	&1.748	&1.082	&-6.0	&81(6)	&1/2\\ \hline
$^{140}\mathsf{La}$	&0.658	&0.322	&-4.0	& 48(2)	&3,4\\ \hline
$^{150}\mathsf{Sm}$	&1.927	&0.773	&3.0	&12(1)	&3,4\\ \hline
$^{156}\mathsf{Gd}$	&1.638	&0.288	&2.4	&23(5)	&1,2\\ \hline
$^{158}\mathsf{Gd}$	&1.517	&0.261	&-0.2	&19(2)	&1,2\\ \hline
$^{160}\mathsf{Tb}$	&0.279	&0.279	&0.12	&23(3)	&1,2\\ \hline
$^{163}\mathsf{Dy}$	&1.055	&0.250	&-3.0	&22(1)	&1/2\\ \hline
$^{164}\mathsf{Dy}$	&1.808	&0.242	&-2.0	&29(1)	&2,3\\ \hline
$^{165}\mathsf{Dy}$	&0.738	&0.184	&-3.6	&53(1)	&1/2\\ \hline
$^{166}\mathsf{Ho}$	&0.522	&0.522	&-1.5	&31(1)	&3,4\\ \hline
$^{168}\mathsf{Er}$	&1.719	&0.995	&-2.3	&27(4)	&3,4\\ \hline
$^{170}\mathsf{Tm}$	&0.715	&0.648	&-1.3	&23(2)	&0,1\\ \hline
$^{174}\mathsf{Yb}$	&1.949	&0.253	&-3.5	&22(1)	&2,3\\ \hline
$^{176}\mathsf{Lu}$	&0.688	&0.595	&-1.8	&44(1)	&3,4\\ \hline
$^{177}\mathsf{Lu}$	&0.854	&0.637	&0.25	&16(1)	&6 $\frac{1}{2}$,7 $\frac{1}{2}$\\ \hline
$^{181}\mathsf{Hf}$	&1.154	&0.332	&-3.1	&52(4)	&1/2\\ \hline
$^{182}\mathsf{Ta}$	&0.480	&0.360	&-2.4	&19(1)	&3,4\\ \hline
$^{183}\mathsf{W}$  &1.471	&0.209	&-4.0	&28(1)	&1/2\\ \hline
$^{184}\mathsf{W}$	&1.431	&0.364	&-2.4	&35(1)	&0,1\\ \hline
$^{185}\mathsf{W}$	&1.106	&1.068	&-0.9	&62(1)	&1/2\\ \hline
$^{187}\mathsf{W}$	&1.083	&0.303	&-2.6	&34(1)	&1/2\\ \hline
$^{188}\mathsf{Os}$	&1.764	&0.633	&-0.2	&59(3)	&0,1\\ \hline
$^{190}\mathsf{Os}$	&1.682	&0.756	&-0.7	&49(3)	&1,2\\ \hline
$^{191}\mathsf{Os}$	&0.815	&0.815	&-3.5	&76(2)	&1/2\\ \hline
$^{192}\mathsf{Ir}$	&0.415	&0.415	&-0.3	&27(6)	&1,2\\ \hline
$^{193}\mathsf{Os}$	&1.288	&0.889	&-3.8	&80(1)	&1/2\\ \hline
$^{196}\mathsf{Pt}$	&1.998	&0.688	&-3.7	&37(5)	&0,1\\ \hline
$^{198}\mathsf{Au}$	&0.528	&0.495	&-5.6	&42(1)	&1,2\\ \hline
$^{200}\mathsf{Hg}$	&1.972	&0.368	&-8.0	&59(2)	&0,1\\ \hline
\end{tabular}
\end{table*}
\section{Results and discussion}
\label{results}
The data on the energies $E_{max}$ of the final level of the cascades and the sums of the experimental intensities are shown in Table \ref{Tab1}. It is seen that, for almost a half of the investigated nuclei, the intensities of measured two-step cascades contain $~ 50\%$ (or more) of the total intensities of all cascade transitions to the final levels. Consequently, for these nuclei the systematic uncertainties of $\rho$ and $\Gamma$ determination are minimized, what means that the fits for the $\rho$ and $\Gamma$ values are the best ones.  

In all calculations for $E_{ex} \leq E_d$ ($E_d$ is the maximal excitation energy of the “discrete” level area) the data on excitation energies and decay modes of low-lying levels from Ref. \cite{nds} were used. 
\subsection{The peculiarities of the fits}
\label{The peculiarities of the fits}
The experimental distributions of the cascade intensities are usually measured in energy intervals with a width of $~ 1$ keV, and can include from 5000 to 10000 channels (Fig. \ref{Fig2}), for each $M$ investigated cascade ($2 \leq M \leq 16$). The basis equations (\ref{2} -- \ref{5}) contain, on the average, $\approx 20$ parameters, which are fitted for all recorded cascades of the investigated nucleus. In practice, for obtaining the fitted parameters, it is reasonable to average the energy intervals of primary transitions and of the excitation energies over $50$ keV.

A solution of the system of Eq. (\ref{1}) is performed by the Monte--Carlo method. The non-linearity of the strongly correlated equations of the system \ref{1} produces an uncertainty of extracting the $\rho$ and $\Gamma$ parameters from $I_{\gamma \gamma}$  intensities. Existence of false local minima of $\chi^2$ is inevitable for the system of nonlinear equations (\ref{1}), and this occurrence puts obstacles in the way of a precise determination of $\rho$ and $\Gamma$ values. A probability to get into the false minimum of $\chi^2$ sometimes amounts to $20\%$. Nevertheless, all accumulated data (see Table \ref{Tab1}) provide new and very important information.

Experimental data on $I_{\gamma\gamma}(E_1)$ are usually obtained with a small total uncertainty  and averaged over $500$ keV energy intervals. The results for $^{156}Gd$ are shown, in more detail, in Figs. \ref{Fig3}, \ref{Fig4} and \ref{Fig5}. The best fits to $I_{\gamma\gamma}(E_1)$, as well as the fitted level densities and strength functions, are compared to corresponding values calculated using the statistical model.  The results and corresponding calculations for the rest of the investigated nuclei are presented in Figs. \ref{Fig6}, \ref{Fig7} and \ref{Fig8} (the cascade intensities), in Figs. \ref{Fig9}, \ref{Fig10} and \ref{Fig11}, (the level densities), in Figs. \ref{Fig12}, \ref{Fig13} and \ref{Fig14} (the radiative strength functions of $E1$-- and $M1$--transitions), in Figs. \ref{Fig15}, \ref{Fig16} and \ref{Fig17} (sums of the strength functions) and in Figs. \ref{Fig18}, \ref{Fig19} and \ref{Fig20} (the ratios of density of vibrational levels to the total level density). The spectra in Figs. \ref{Fig3}, \ref{Fig6}, \ref{Fig7} and \ref{Fig8} were calculated using functions which are shown as solid lines in Figs. \ref{Fig15}, \ref{Fig16} and \ref{Fig17}. 
\subsection{The resulted parameters}
\label{resPar} 
Various shapes of the $I_{\gamma\gamma}$ distributions for different nuclei (Figs. \ref{Fig3}, \ref{Fig6}, \ref{Fig7} and \ref{Fig8}) are most likely determined by a diverse structure of the wave functions of exited levels. In a similar manner, for example, the very strong dependence of neutron strength functions on nuclear mass \cite{Bohr} or the dependence of spectroscopic factors of reactions (d, p) and (d, t) on the locations of low-lying levels (relative to the Fermi-surface) \cite{Bond} are explained.  

The level densities in Figs. \ref{Fig4}, \ref{Fig9}, \ref{Fig10} and \ref{Fig11} demonstrate that, if the shell inhomogeneities of single-particle spectra are taken into account, the single-particle densities noticeably reduce in comparison to the ones calculated with the hypothesis $a=const$. At that, level densities obtained in our model slightly change when the shell corrections are taken into account. Thus, in Figs. \ref{Fig4}, \ref{Fig9}, \ref{Fig10} and \ref{Fig11} the curves that describe the calculated single-particle density (using Eq. \ref{6}) and the ones for fitted level density, for all investigated nuclei, became closer to each other.

The main source of large fluctuations of radiative strength functions (see Figs. \ref{Fig5},\ref{Fig12},\ref{Fig13} and \ref{Fig14}) is their anti-correlation with the level density in every energy range. Average sums of the strength functions of $E1$-- and $M1$--transitions for $E_1 = 520$ keV are $0.80(8)$, $2.1(2)$ and $2.5(3)\cdot 10^{-10} MeV^{-3}$ for even-even, even-odd and odd-odd nuclei, respectively. Thus, the summation noticeably reduces the above-mentioned scatter and allows us to assert that the sum strength function decreases with reducing the energy of the primary transition.

The contributions of the levels of vibrational type in the total level density (Figs. \ref{Fig18}, \ref{Fig19} and \ref{Fig20}) for all nuclei decrease near the $U_l$ points. For a majority of the nuclei the part of vibrational levels below $B_n$ is about $40 \%$, which does not contradict the analysis of distributions of the total radiative widths above $B_n$ \cite{Sukho8}. Calculations of the distributions of random deviations for the total radiative widths of s-resonances, executed in Ref. \cite{Sukho8}, showed that there is a superposition, at least, of four distributions with different average $\left\langle \Gamma_{\gamma}\right\rangle$. 

When investigating the gamma-decay process, problem of the description of special points (of the breaking of the Cooper pairs) appears. As anti-correlation between $\rho$ and $\Gamma$ values can be different to a greater or lesser extent in all excitation energy range, it can be maximal just in the points of breaking the Cooper pairs. There is a noticeable dependence of the resulted strength functions on the local jumps in the level density. At that, as it was already pointed out, in order to prevent a contradiction between the data of the two-step experiment and the data from one-step experiment, it is necessary to take into account the connection between $\rho$ and $\Gamma$ values.  We have made an effort to investigate such anti-correlation by means of multiplication of the phenomenological expression (\ref{5}) for the strength function by $\rho_{mod}/\rho_{exp}$ ratio, which inserts an additional fitted correlation. Here $\rho_{exp}$ is taken from the best fit during solving the system (\ref{1}) and ρmod is taken from the back-shifted Fermi-gas model \cite{Dilg}. The function $\rho_{mod}$ represents smoothed density for levels of fermion type and describes both a neutron resonance density and the cumulative sum of known nuclear levels at $E_{ex}\leq ≤ E_d$ ($E_d$ was taken from Ref. \cite{nds}). The limiting condition $1\leq \rho_{mod}/\rho_{exp}\leq 10$ from Ref. \cite{Sukho2,Mitsy} was implemented in this analysis.

Introducing the coefficient $\rho_{mod}/\rho_{exp}$ to the phenomenological formula for the strength function, which makes the residual anti-correlation of the fitted $\rho$ and $\Gamma$ demonstrable, was done for testing the influence on the shape of the strength functions of the step-like structure of fitted level density distribution. Simultaneously, it was a test of the invariability of this step-like structure. 

\subsection{The Cooper pairs breaking thresholds }
\label{thresholds}
\begin{figure}[h] 
\includegraphics[scale=0.21]{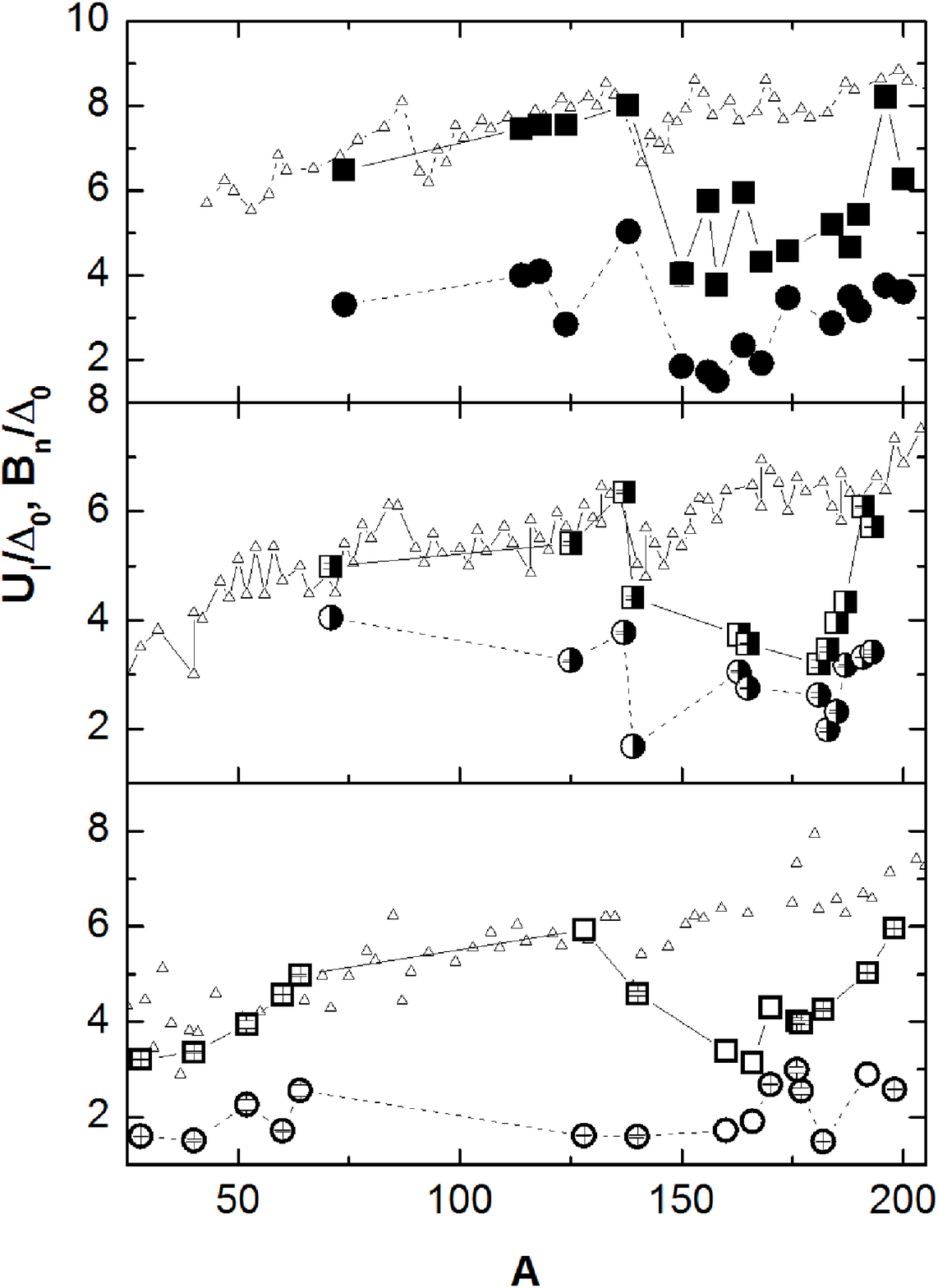}
\centering
\caption{\label{Fig21}
Dependence on the nuclear mass $A$ of the ratios $U_l/\Delta_0$, of break-up thresholds to the average pairing energy of the last nucleon, for the second (points) and the third (squares) Cooper pairs. Full points are even-even, half-open points are even-odd and open points are odd-odd compound nuclei. Triangles show the mass dependence of $B_n/\Delta_0$ ratio.
}
\end{figure}

The connection between the shape of the investigated nucleus and the breaking thresholds, which was established for the first time in our prior analysis (Ref. \cite{Mitsy,Sukho6,Sukho7}), was confirmed again in the present analysis. As the breaking thresholds differ for nuclei with various nucleon parities and depend on the average pairing energy $(\Delta_0)$ of the last nucleon, the mass dependencies for the ratios of the break up thresholds of the second and the third Cooper pairs to $\Delta_0$, as well as the mass dependence of the binding energy to $\Delta_0$, are presented in Fig. \ref{Fig21}. As it can be seen in Fig. \ref{Fig21}, there is a noticeable difference in $U_2/\Delta_0$ and $U_3/\Delta_0$ ratios for spherical and deformed nuclei in contrast to $B_n/\Delta_0$. 
\begin{figure}[ht] 
\includegraphics[scale=0.2]{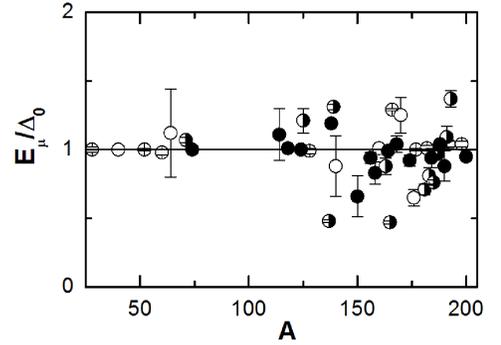}
\centering
\caption{\label{Fig22}
Dependence of $E_{\mu}$ and $E_{\nu}$ model parameters on the nuclear mass $A$. Full points are even-even, half-open are even-odd and open points are odd-odd nuclei.}
\end{figure}
\begin{figure}[h] 
\includegraphics[scale=0.2]{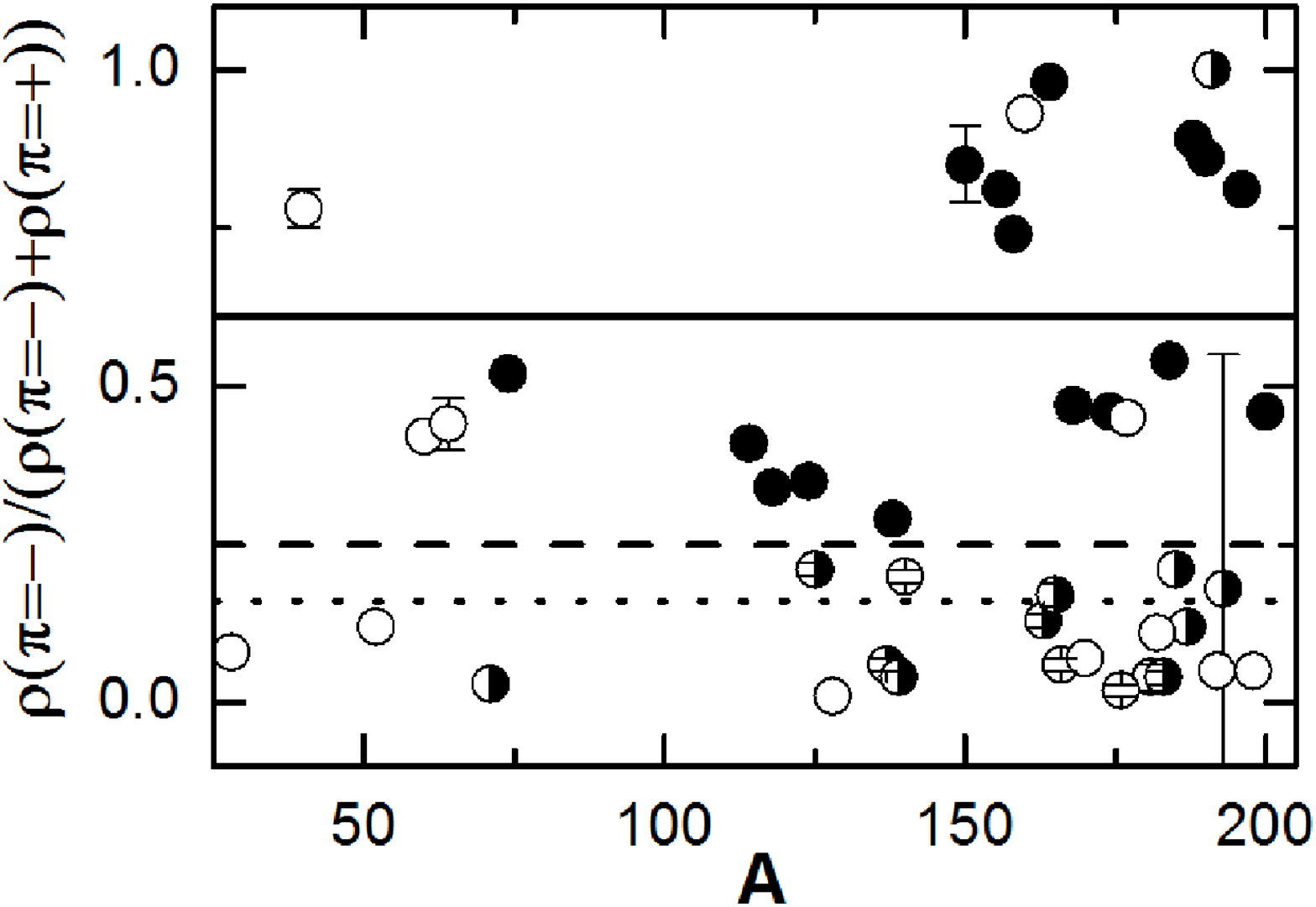}
\centering
\caption{\label{Fig23}
Mass dependence of the ratio of the level density with negative parity to the total level density at the upper energy border of the level discrete region ($E_d$) and their averages for even-even nuclei (solid lines), even-odd (dashed lines) and odd-odd nuclei (dotted lines). Full points are even-even, half-open points are even-odd and open points are odd-odd compound nuclei. 
}
\end{figure}
\subsection{About level parity}
\label{level}
For determination of the part $r = \rho(\pi−)/(\rho(\pi−)+\rho(\pi+))$ of levels $\rho(\pi−)$ with negative parity, a linear extrapolation for $r$ value was applied in the $E_d\leq E_{ex} \leq B_n$ energy interval. At that, in the $B_n$ point we use generally accepted assumption, that $\rho(\pi−) = 0.5\cdot(\rho(\pi−) + \rho(\pi+))$, $\rho(\pi−)$ value in this energy point was fixed, and at the $E_d$ energy the $\rho(\pi−)$ value varied. 

The calculated ratios of density of the levels with negative parity to the total level density are shown in Fig. \ref{Fig22}. The averages of these ratios are $0.61(22)$, $0.25(28)$ and $0.16(16)$ for even-even, even-odd and odd-odd nuclei, respectively (and for odd-even $^{177}Lu$ it is $0.65(1)$). Hence, the behavior of the gamma-decay process is different for nuclei of various nucleon parities. 
%
\section{Possible experiments for a study of superfluidity}
\label{Possible experiments for a study of superfluidity}
Experiments on recording the cascades of two gamma-transitions of radiative capture of thermal neutrons were carried out in Dubna (Russia), Riga (Latvia), Rez (Czech Republic) and Dalat (Vietnam). Unfortunately, gamma-ray cascades at thermal neutron capture allow the determination of $\rho$ and $\Gamma$ only in a fixed area of nuclear excitations, for a certain spin interval and for a given parity of the decaying resonance (Table \ref{Tab1}). 

Until now, in all analysis, a nucleus was usually imagined as a statistical system. The real uncertainty of this nuclear model is yet unknown, therefore new experiments (e.g. as in Ref. \cite{Sukho2}) are needed. Such an experiment can be carried out not only at the beam of thermal and resonance neutrons, but at any accelerators of charged particles, if the scatter of energies of excited levels $\lambda$ in the target and the energy resolution of the HPGe -- detectors are comparable. 

The best approach to study the cascades of gamma-transitions of decaying levels excited by gamma-quanta can be realized at any source of gamma-radiation (e.g. ELBE \cite{Maki} or S-DALINAC \cite{Ozel}) with a fixed energy. At fixed energy $E_{max}$ of the gamma beam, it is possible to apply the model of Ref. \cite{Sukho2} in the interval of excitation energies of the decaying levels from $E_{max}$ to $E_{max} - 511$ keV. It would allow for the process of the cascade decay to become clearer. 

The background conditions during cascade recordings for a beam of gamma-quanta are essentially better than for the neutron beams. For experiments of the types seen in Ref. \cite{Maki,Ozel} a singular requirement is needed – detectors must be placed closely to the target, in a back hemisphere relatively to it. In such an experiment it is possible to determine separately the radiative strength functions for gamma-transitions both to the ground state of the target-nucleus, and to its’ excited levels. The information content of such an experiment will exceed the results of $(n, 2\gamma)$ reaction investigation at least ten times.

Unlike the cascades of gamma-transitions, the cascades with nucleon emission provide significant statistical improvements due to a high efficiency of the recording charged products of the reaction. Mathematically, a spectrum of primary gamma-transitions of decaying levels below the emission threshold for nucleon products of the reaction and a spectrum of evaporated nucleons (light nuclei) above the binding energy are identical. Therefore, the analysis of a cascade of evaporated nucleon and gamma-quanta is similar to the analysis of the cascade of two gamma-transitions. But the intensity of a cascade of nucleon and gamma- quantum to low-lying levels can be strongly dependent on the orbital moment of the evaporated nucleon. 

\section{Conclusion}
\label{Conclusion}
In order to obtain reliable values of $\rho$ and $\Gamma$ values, an effective practical model, that takes into account the interaction of fermion and boson components in the nuclear matter, is used.

A necessity of taking into account the corrections for the shell inhomogeneities of single-particle spectrum on the level density was demonstrated when comparing the parameters obtained in two different conditions: at the constant density of single-particle states near Fermi-surfaceand at $g \neq const$. The results obtained when using the shell corrections became closer to the existing representations. Nevertheless, we cannot describe the cascade intensities without taking into account a strong influence of the nuclear superfluidity on the gamma-decay process. 

The data on Cooper pair break-up energies, obtained with a high accuracy, are sufficient to conclude that the dynamics of interaction between superfluid and normal phases of a nucleus depends on its’ shape.

Our model allows for a separate determination of the density of vibrational levels between the breaking thresholds of the Cooper pairs. 

The common result for nuclei, in two-step gamma-decay, with different parity of nucleons is a decrease in the sum $k$ of radiative strength functions when the energy of primary transitions descends. When one analyzes the set of investigated nuclei, the average sums are almost equal for the even-odd and odd-odd nuclei, while $k$ values are two times smaller for the even-even nuclei.

Unfortunately, an existence of the sources of uncertainties of the sought $\rho$ and $\Gamma$ functions is a fundamental problem, and it is inevitable for any nuclear model used for experimental data analysis and for predictions of the spectra and cross sections. There are also fluctuations of the intensities of gamma-transitions in different nuclei, which have a contribution to the systematical error. Nevertheless, the practical model showed one possibility to describe the data of the two-step experiments with the accuracy that exceeds the statistical one. 

\begin{acknowledgments}
We wish to express our gratitude to Dr Petar Mali for helpful advice and discussions.

\end{acknowledgments}



\end{document}